\documentclass[journal,draftcls,onecolumn,12pt,twoside]{IEEEtranTCOM}
\normalsize

\ifCLASSINFOpdf
\else
\fi

\DeclareMathAlphabet{\mathsfbr}{OT1}{cmss}{m}{n}
\SetMathAlphabet{\mathsfbr}{bold}{OT1}{cmss}{bx}{n}
\DeclareRobustCommand{\msf}[1]{%
  \ifcat\noexpand#1\relax\msfgreek{#1}\else\mathsfbr{#1}\fi
}

\makeatletter
\newcommand{\msfgreek}[1]{\csname s\expandafter\@gobble\string#1\endcsname}
\makeatother

\DeclareFontEncoding{LGR}{}{} 
\DeclareSymbolFont{sfgreek}{LGR}{cmss}{m}{n}
\SetSymbolFont{sfgreek}{bold}{LGR}{cmss}{bx}{n}
\DeclareMathSymbol{\salpha}{\mathord}{sfgreek}{`a}
\DeclareMathSymbol{\sbeta}{\mathord}{sfgreek}{`b}
\DeclareMathSymbol{\sgamma}{\mathord}{sfgreek}{`g}
\DeclareMathSymbol{\sdelta}{\mathord}{sfgreek}{`d}
\DeclareMathSymbol{\sepsilon}{\mathord}{sfgreek}{`e}
\DeclareMathSymbol{\szeta}{\mathord}{sfgreek}{`z}
\DeclareMathSymbol{\seta}{\mathord}{sfgreek}{`h}
\DeclareMathSymbol{\stheta}{\mathord}{sfgreek}{`j}
\DeclareMathSymbol{\siota}{\mathord}{sfgreek}{`i}
\DeclareMathSymbol{\skappa}{\mathord}{sfgreek}{`k}
\DeclareMathSymbol{\slambda}{\mathord}{sfgreek}{`l}
\DeclareMathSymbol{\smu}{\mathord}{sfgreek}{`m}
\DeclareMathSymbol{\snu}{\mathord}{sfgreek}{`n}
\DeclareMathSymbol{\sxi}{\mathord}{sfgreek}{`x}
\DeclareMathSymbol{\somicron}{\mathord}{sfgreek}{`o}
\DeclareMathSymbol{\spi}{\mathord}{sfgreek}{`p}
\DeclareMathSymbol{\srho}{\mathord}{sfgreek}{`r}
\DeclareMathSymbol{\ssigma}{\mathord}{sfgreek}{`s}
\DeclareMathSymbol{\stau}{\mathord}{sfgreek}{`t}
\DeclareMathSymbol{\supsilon}{\mathord}{sfgreek}{`u}
\DeclareMathSymbol{\sphi}{\mathord}{sfgreek}{`f}
\DeclareMathSymbol{\schi}{\mathord}{sfgreek}{`q}
\DeclareMathSymbol{\spsi}{\mathord}{sfgreek}{`y}
\DeclareMathSymbol{\somega}{\mathord}{sfgreek}{`w}

\DeclareMathSymbol{\svarsigma}{\mathord}{sfgreek}{`c}

\DeclareMathSymbol{\sGamma}{\mathalpha}{sfgreek}{`G}
\DeclareMathSymbol{\sDelta}{\mathalpha}{sfgreek}{`D}
\DeclareMathSymbol{\sTheta}{\mathalpha}{sfgreek}{`J}
\DeclareMathSymbol{\sLambda}{\mathalpha}{sfgreek}{`L}
\DeclareMathSymbol{\sXi}{\mathalpha}{sfgreek}{`X}
\DeclareMathSymbol{\sPi}{\mathalpha}{sfgreek}{`P}
\DeclareMathSymbol{\sSigma}{\mathalpha}{sfgreek}{`S}
\DeclareMathSymbol{\sUpsilon}{\mathalpha}{sfgreek}{`U}
\DeclareMathSymbol{\sPhi}{\mathalpha}{sfgreek}{`F}
\DeclareMathSymbol{\sPsi}{\mathalpha}{sfgreek}{`Y}
\DeclareMathSymbol{\sOmega}{\mathalpha}{sfgreek}{`W}

\DeclareRobustCommand{\mcal}[1]{%
  \ifcat\noexpand#1\relax\mathnormal{#1}\else\cal{#1}\fi
}
\DeclareRobustCommand{\BM}[1]{%
  \ifcat\noexpand#1\relax\bm{\boldUppercaseItalicGreek{#1}}\else\bm{#1}\fi
}
\makeatletter
\newcommand{\boldUppercaseItalicGreek}[1]{\csname var\expandafter\@gobble\string#1\endcsname}
\makeatother

\newcommand{\V}[1]{\bm{#1}} 
\newcommand{\M}[1]{\BM{#1}} 
\newcommand{\Set}[1]{{\mcal{#1}}} 

\newcommand{\E}[1]{\mathbb{E}\left\{#1\right\}}

\newcommand{\avg}[1]{\overline{\left\{#1\right\}}}

\renewcommand{\IEEEQED}{\IEEEQEDopen}

\usepackage{bm}
\usepackage{color, courier}
\usepackage{psfrag}
\usepackage{acronym}
\usepackage{amsmath}
\usepackage{amssymb}
\usepackage{amsfonts}
\usepackage[colorlinks = true, allcolors = black]{hyperref}
\usepackage{latexsym}
\usepackage{mathrsfs}
\usepackage{epstopdf}
\usepackage{algorithm}
\usepackage{mathtools}
\usepackage[caption=false, font=scriptsize]{subfig}
\usepackage{algorithmic}
\usepackage{relsize}
\usepackage{comment}

\graphicspath{{figures/}}

\DeclareMathOperator*{\argmax}{arg\,max}

\newcommand{\st}{\operatorname{s.t.}\,}

\newtheorem{definition}{Definition}
\newtheorem{lemma}{Lemma}
\newtheorem{proposition}{Proposition}
\newtheorem{theorem}{Theorem}

\newtheorem{remark}{Remark}

\definecolor{green}{rgb}{0, 0.5, 0}
\definecolor{pink}{rgb}{1, 0, 1}

\newcommand{\blue}[1]{{\color{blue}#1}}

\acrodef{agi}[AgI]{augmented information}
\acrodef{ldp}[LDP]{Lyapunov drift-plus-penalty}
\acrodef{lp}[LP]{linear programming}
\acrodef{ap}[AP]{access point}
\acrodef{pdf}[pdf]{probability density function}
\acrodef{ue}[UE]{user equipment}
\acrodef{es}[ES]{edge server}
\acrodef{sfc}[SFC]{service function chain}
\acrodef{bp}[BP]{back-pressure}
\acrodef{mec}[MEC]{mobile edge computing}

\acrodef{nfv}[NFV]{network function virtualization}
\acrodef{sdn}[SDN]{software defined networking}

\acrodef{wrt}[w.r.t.]{with respect to}
\acrodef{wlog}[w.l.o.g.]{Without loss of generality}

\acrodef{umw}[UMW]{universal max-weight}

\begin{document}

%
\title{
    Decentralized Control of Distributed Cloud Networks with Generalized Network Flows
}

%
%
%

\author{
    Yang~Cai,~\IEEEmembership{Graduate~Student~Member,~IEEE},
    Jaime~Llorca,~\IEEEmembership{Member,~IEEE},
    Antonia~M.~Tulino,~\IEEEmembership{Fellow,~IEEE}, and
    Andreas~F.~Molisch,~\IEEEmembership{Fellow,~IEEE}%
    \thanks{Part of this work was presented at the 2021 IEEE ICC \cite{cai2021multicast}.
    A condensed version of this paper has been submitted to the IEEE Transactions on Communications \cite{cai2022multicast_tcom}. 
    }%
    \thanks{Y. Cai and A. F. Molisch are with the Department of Electrical and Computer Engineering, University of Southern California, Los Angeles, CA 90089, USA (e-mail: yangcai@usc.edu; molisch@usc.edu).}%
    \thanks{J. Llorca is with the Electrical and Computer Engineering Department, New York University, Brooklyn, NY 11201 USA (e-mail: jllorca@nyu.edu).}%
    \thanks{A. M. Tulino is with the Electrical and Computer Engineering Department, New York University, Brooklyn, NY 11201 USA, and also with the Department of Electrical Engineering, University\`{a} degli Studi di Napoli Federico II, Naples 80138, Italy (e-mail: atulino@nyu.edu; antoniamaria.tulino@unina.it).}%
    \thanks{This work was supported by the National Science Foundation (NSF) under CNS-1816699 and RINGS-2148315. 
    }%
}

%
%

\markboth{IEEE Transactions on Communications}%
{Submitted paper}
%



\maketitle

\vspace{-50pt}

\begin{abstract}
\vspace{-10pt}
Emerging distributed cloud architectures, e.g., fog and mobile edge computing, are playing an increasingly important role in the efficient delivery of real-time stream-processing applications (also referred to as augmented information services), such as industrial automation and metaverse experiences (e.g., extended reality, immersive gaming). While such applications require processed streams to be shared and simultaneously consumed by multiple users/devices, existing technologies lack efficient mechanisms to deal with their inherent multicast nature, leading to unnecessary traffic redundancy and network congestion. In this paper, we establish a unified framework for distributed cloud network control with generalized (mixed-cast) traffic flows that allows optimizing the distributed execution of the required packet processing, forwarding, and replication operations. We first characterize the {\em enlarged} multicast network stability region under the new control framework (with respect to its unicast counterpart). We then design a novel queuing system that allows scheduling data packets according to their current destination sets, and leverage Lyapunov drift-plus-penalty control theory to develop the {\em first fully decentralized, throughput- and cost-optimal algorithm for multicast flow control}. Numerical experiments validate analytical results and demonstrate the performance gain of the proposed design over existing 
network control policies.
\end{abstract}

\vspace{-10pt}
\begin{IEEEkeywords}
\vspace{-10pt}
Distributed computing, 
routing, flow control, distributed control, cost optimal control.
\end{IEEEkeywords}

%
\IEEEpeerreviewmaketitle

\acresetall

\vspace{-10pt}

\section{Introduction}
\label{sec:introduction}

\IEEEPARstart{A}{pplications} receiving the most recent attention, such as system automation (e.g., smart homes/factories/cities, self-driving cars) and metaverse experiences (e.g., multiplayer gaming, immersive video, virtual/augmented reality) are characterized by intensive resource consumption and real-time interactive requirements, where multiple users/devices simultaneously consume information that results from the real-time processing of a variety of live multimedia streams \cite{cai2022metaverse}. We refer to this general class of 
services as \ac{agi} services.

The high computation and low latency requirements of \ac{agi} services are fueling the evolution of network architectures toward widespread deployments of increasingly distributed computation resources, leading to what is referred to as {\em distributed cloud/computing networks}, including fog and \ac{mec} \cite{Wel:B16}. In contrast to traditional architectures, where there is a clear separation between data processing at centralized data centers and data transmission between remote data centers and end users, distributed cloud networks are evolving toward tightly integrated compute-communication systems, where the ubiquitous availability of computation resources enables reduced access delays and energy consumption, thus providing better support for next-generation delay-sensitive, compute-intensive \ac{agi} services.

In addition, the increasing amount of real-time multi-user interactions present in \ac{agi} services, where media streams can be shared and simultaneously consumed by a large number of users/devices, is creating the need to efficiently support multicast traffic. While, in general, one can describe four types of network flows, i.e., unicast (packets intended for a unique destination), multicast (packets intended for multiple destinations), broadcast (packets intended for all destinations), and anycast (packets intended for any node in a given group), we focus our attention on multicast flows as the most general class, given that the other three types either become special cases of or can be transformed into multicast flows.

To maximize the benefit of distributed cloud networks for the delivery of next-generation multicast AgI services, two fundamental flow control problems need to be addressed:
{\em \begin{itemize}
	\item {\bf \em packet processing:} where to process data packets by 
	the required service functions 
	and how much computation resource to allocate
	\item {\bf \em packet forwarding and replication:} how to route data packets through the required sequence of service functions, where to replicate data packets in order to satisfy the demand from multiple destinations, and how much network resource to allocate
\end{itemize}}

In addition, the above processing, routing (including forwarding and replication), and resource allocation problems must be addressed in an online manner, in response to unknown time-varying network conditions and service demands.

\subsection{Related Work}

With the advent of software defined networking (SDN) and network function virtualization (NFV), network services can be deployed as a sequence of software functions or \acp{sfc} that can be flexibly interconnected and elastically executed at distributed cloud locations \cite{Wel:B16}. A number of studies have investigated the problem of joint \ac{sfc} placement and routing with the objective of either maximizing accepted service requests \cite{huang2021throughput,yue2021throughput}, or minimizing overall resource cost \cite{BarLloTulRam:C15,barcelo2016IoT,Feng_et_al_2017_Infocom}. Nonetheless, these solutions exhibit two main limitations: first, the problem is formulated as a {\em static} optimization problem without considering the dynamic nature of service requests, a critical aspect in next-generation \ac{agi} services; second, due to the combinatorial nature of the problem, the corresponding formulations typically take the form of (NP-hard) mixed integer linear programs, and either heuristic solutions or loose approximation algorithms are developed, compromising the quality of the resulting solution.

More recently, another line of work has addressed the \ac{sfc} optimization problem in {\em dynamic} scenarios, where one needs to make joint packet processing and routing decisions in an online manner \cite{kim2021dynamic, FenLloTulMol:J18a,FenLloTulMol:J18b,zhang2021multicast}.
The work in \cite{kim2021dynamic} leverages Lagrange duality and saddle point theory to design an iterative algorithm that employs global candidate path information to optimize service chain multi-path routing and associated sending rates.
The works in \cite{FenLloTulMol:J18a,FenLloTulMol:J18b} employ a generalized cloud network flow model that allows joint control of processing and transmission flows. The work in \cite{zhang2021multicast} shows that the cloud network flow control problem (involving processing and transmission decisions) can be reduced to a packet routing problem on a {\em cloud layered graph} that includes extra edges to characterize the computation operations (i.e., data streams pushed through these edges are interpreted as being processed by a service function). By this transformation, control policies designed for packet routing can be extended to address cloud network control problems.

{\em Dynamic unicast routing} is a long-explored problem, with a number of existing algorithms known to maximize network throughput with bounded average delay. In particular, the \ac{bp} algorithm \cite{TasEph:J92} is a well-known approach for throughput-optimal routing that leverages Lyapunov drift control theory to steer data packets based on the {\em pressure} difference (differential queue backlog) between neighbor nodes. In addition, the \ac{ldp} control approach \cite{Nee:B10} extends the \ac{bp} algorithm to also minimize network operational cost (e.g., energy expenditure), while preserving throughput-optimality. However, both \ac{bp} and \ac{ldp} approaches can suffer from poor average delay performance, especially in low congestion scenarios, where packets can follow unnecessarily long, and sometimes even cyclic, paths for delivery \cite{BuiSriSto:C09}. To address this problem, \cite{YinShaRedLiu:J11} combines \ac{bp} and hop-distance based shortest-path routing, which can effectively reduce the average delay.

Going beyond unicast traffic and addressing the design of optimal routing policies for multicast traffic is a much more challenging problem, as the need for packet replications violates the flow conservation law \cite{zhang2021multicast}. Despite the large body of existing works on this topic \cite{8116620}, {\em throughput-optimal least-cost multicast packet routing} remains an open problem. Even under {\em static} arrivals, the Steiner tree problem, which aims to find the multicast tree (a tree that covers all destinations) with minimum weight, is known to be NP-hard \cite{kleinberg2006algorithm}. Many heuristic approaches have been developed to address this problem, such as the Extended Dijkstra's Shortest Path Algorithm (EDSPA) \cite{ananta2014multicasting}, which delivers multicast packets along a tree formed by the union of the shortest paths from the source to all destinations. However, in addition to their heuristic nature, packets are delivered along fixed paths under these policies, lacking dynamic exploration of route and processing diversity.  Considering {\em dynamic} arrivals becomes a further challenge that requires additional attention. A centralized dynamic packet routing and scheduling algorithm, UMW, was proposed in \cite{SinMod:J18}, shown to achieve optimal throughput with mixed-cast network flows. Nonetheless, this design exhibits two limitations:
(i) it makes centralized decisions based on global queuing information, incurring additional communication overhead, 
(ii) it leaves out operational cost minimization, 
an important aspect in modern elastic network environments.

\subsection{Contributions}

The goal of this work is to develop {\em decentralized} control policies for distributed cloud network flow problems dealing with mixed-cast \ac{agi} services. We establish a new multicast control framework 
that guides the creation of copies of data packets as they travel toward their corresponding destinations. Compared to the state-of-the-art unicast approach \cite{FenLloTulMol:J18a} that ``creates one copy for each destination of a multicast packet upon arrival, and treats them as individual unicast packets'', the proposed policy employs a {\em joint forwarding and replication} strategy that (i) eliminates redundant transmissions along network links common to multiple copies' routes, and (ii) reduces computation operations by driving computation before replication when beneficial. 


The proposed approach is based on
a novel multicast queuing system that allows formalizing the {\em packet replication operation}, which defines ``where to create copies'' and ``how to assign destinations to resulting copies'', as a network control primitive. Under the proposed queuing system, each packet is labeled and queued according to its {\em replication status}, which keeps track of its current destination set. The packet replication operation is then specified by the partition of the destination set of a given packet into the destination sets of each resulting copy. We finally devise a fully decentralized packet processing, forwarding, and replication policy that attains optimal throughput and cost performance (see Theorem \ref{thm:tradeoff} for details), as well as a variant achieving sub-optimal performance with polynomial complexity.

Our contributions can be summarized as follows:
\begin{enumerate}
  
  \item We establish a novel queuing system to accommodate packets according to their replication status that allows supporting packet processing, routing, and replication operations as necessary cloud network control primitives for the delivery of mixed-cast AgI services.
  
  \item We characterize the {\em enlarged} multicast network stability region obtained by including packet replication as an additional control primitive, and quantify the resulting gain \ac{wrt} its unicast counterpart.
  
  \item We devise GDCNC, the first fully decentralized, throughput- and cost-optimal algorithm for distributed cloud network control with mixed-cast network flows.
  
  \item We design GDCNC-R, a computational-efficient policy achieving sub-optimal performance with polynomial complexity by focusing on a subset of effective replication operations.
  
  \item We conduct extensive numerical experiments that support analytical results and demonstrate the performance benefits of the proposed design for the delivery of mixed-cast AgI services.
  
\end{enumerate}

{\bf Organization:}
In Section \ref{sec:system_model}, we introduce the model for the ``multicast packet routing'' problem. In Section \ref{sec:policy_space}, we define the policy space and present a characterization for the multicast network stability region. Section \ref{sec:queuing_system} describes the multicast queuing system and defines the problem formulation. In Section \ref{sec:exact_solution}, we devise the GDCNC control policy and analyze its performance, which further motivates the design of GDCNC-R in Section \ref{sec:apx_solution}. Extensions to the proposed design are discussed in Section \ref{sec:ext}. Section \ref{sec:experiments} presents the numerical results, and conclusions are drawn in Section \ref{sec:conclusion}.

{\bf Notation:}
Let $\{0, 1\}^{D}$ denote the set of all $D$-dimensional binary vectors. We use $\V{0}$ to denote the zero vector, $\V{1}$ the all-ones vector, and $b_k$ the vector with only the $k$th entry equal to $1$ (and $0$'s elsewhere). Given a binary vector $q\in\{0, 1\}^{D}$, $\bar{q} = \V{1} - q$ denotes its complement and $2^{q} \triangleq \{ s: s_k = q_k u_k, u\in \{0, 1\}^D \}$ its power set. The inner product of vector $x$ and $y$ is denoted by $\langle x, y \rangle$. $\E{z}$ is the expectation of random variable $z$, and $\avg{ z(t) } \triangleq \lim_{T\to\infty} (1/T) \sum_{t=0}^{T-1}{ z(t) }$ the time average of random process $\{z(t) : t\geq 0 \}$. $\mathbb{I}\{ \mathscr{A} \}$ denotes the indicator function (equal to $1$ if event $\mathscr{A}$ is true, and $0$ otherwise), and $|\Set{A}|$ the cardinality of set $\Set{A}$.

\section{System Model}
\label{sec:system_model}

\subsection{Cloud Layered Graph}

The ultimate goal of this work is to design decentralized control policies for distributed cloud networks equipped with computation resources (e.g., cloud servers, edge/fog computing nodes, etc.) able to host \ac{agi} service functions and execute corresponding computation tasks. 

While in traditional packet routing problems, each node treats its neighbor nodes as outgoing interfaces over which packets can be scheduled for transmission, a key step to address the \ac{agi} service delivery problem, involving packet processing and routing decisions, is to treat co-located 
computing devices as an additional interface over which packets can be scheduled for processing \cite{FenLloTulMol:J18a}. As illustrated in \cite{zhang2021multicast}, the \ac{agi} service control problem can be transformed into a packet routing problem on a {\em layered graph} where cross-layer edges represent computation resources.

Motivated by such a connection and for ease of exposition, in this paper, without loss of generality, we illustrate the developed approach focusing on the {\em single-commodity, least-cost multicast packet routing} problem. We remark that (i) the optimal decentralized multicast control problem 
has been open up to now even in traditional communication networks, and (ii) the extension to distributed cloud network control is presented in Section \ref{sec:AgI}.

\subsection{Network Model}
\label{sec:routing_net}

The considered packet routing network is modeled by a directed graph $\Set{G} = (\Set{V}, \Set{E})$, where each edge $(i, j) \in \Set{E}$ represents a network link supporting data transmission from node $i$ to $j$, and $\delta_i^-$ and $\delta_i^+$ denote the incoming and outgoing neighbor sets of node $i$, respectively.

Time is slotted. For each link $(i, j) \in \Set{E}$, we define: (i) {\em transmission capacity} $C_{ij}$ as the maximum number of data units (e.g., packets) that can be transmitted in one time slot, and (ii) {\em unit transmission cost} $e_{ij}$ as the cost (e.g., power consumption) incurred in transmitting one unit of data in one time slot.
We note that a linear cost model is used here for ease of exposition, while extensions to a non-linear cost model characterizing power expenditure in a practical wireless scenario is studied in Section \ref{sec:mec}.

We emphasize that in the layered graph, cross-layer edges represent data processing, i.e., data streams pushed through these edges are interpreted as being processed by corresponding service functions, and the capacity and cost of these edges represent the {\em processing capacity} and {\em unit processing cost} of the associated computation resources (e.g., cloud/edge servers).

\subsection{Arrival Model}
\vspace{-2pt}

We focus on a multicast application (the extension to multiple applications is straightforward, and details are given in 
Appendix \ref{apdx:AgI})
where each incoming packet is associated with a destination set $\Set{D} = \{d_1, \cdots, d_D \} \subset \Set{V}$, with $d_k$ denoting the $k$-th destination and $D = |\Set{D}|$ the destination set size. At least one copy of the incoming packet must be delivered to every destination in $\Set{D}$. Importantly, we assume that delivering multiple copies of the same packet (containing the same content) to a given destination does not increase network throughput.

Multicast packets originate at the application source nodes $\Set{S} \subset \Set{V}\setminus \Set{D}$. Let $a_i(t)$ denote the number of exogenous packets arriving at node $i$ at time $t$, with $a_i(t)=0, \forall i\notin \Set{S}$. We assume that the arrival process is i.i.d. over time, with mean arrival rate $\lambda_i \triangleq \mathbb{E}\big\{a_i(t)\big\}$ and bounded instantaneous arrivals; the corresponding vectors are denoted by $\V{a}(t) = \big\{ a_i(t): i\in \Set{V} \big\}$ and $\V{\lambda}$. 

\begin{remark}
By properly defining the destination set $\Set{D}$, the above model can capture all four network flow types:
(i) unicast and broadcast flows are special cases of a multicast flow, defined by setting $\Set{D} = \{d\}$ and $\Set{D} = \Set{V}$, respectively;
and (ii) an anycast flow can be transformed into a unicast flow by creating a {\em super destination node} connected to all the candidate destinations \cite{SinMod:J18}. Therefore, it suffices to focus on the multicast flow to derive a general solution for the mixed-cast flow control problem.
\end{remark}

\subsection{In-network Packet Replication}
\label{sec:duplication}

We now formalize the most important operation for multicast packet routing, namely {\em in-network packet replication}.

\subsubsection{Replication Status}

We assume that each packet is associated with a label that indicates its {\em current} destination set, i.e., the set of destinations to which a copy of the packet must still be delivered, formally defined as follows.

\begin{definition}
For a given packet, the {\em replication status} $q  = [q_1, \cdots, q_D]$ is a $D$-dimensional binary vector where the $k$-th entry ($k=1,\cdots, D$) is set to $q_k = 1$ if destination $d_k\in \Set{D}$ belongs to its current destination set, and to $q_k = 0$ otherwise.
\end{definition}

Three important cases follow:
(i) $q = \V{1}$ indicates the status of every newly arriving packet prior to any replication operation, with the entire destination set $\Set{D}$ as their current destination set;
(ii) $q = b_k$ describes a packet with one destination $d_k$, which behaves like a unicast packet;
(iii) $q = 0$ describes a packet without any destination, which is removed from the system immediately.

\subsubsection{Packet Replication and Coverage Constraint}

A replication operation creates copies of a packet and assigns a new destination set, which must be a subset of the original packet's destination set, to each resulting copy. Let $q\in \{0, 1\}^D$ denote the replication status of the original packet; then, the set of all possible replication status of the copies is given by its power set $2^q$. To ensure the delivery of a packet to all of its destinations, we impose the {\bf Coverage} constraint on the replication operation: each destination node of the original packet must be present in the destination set of {\em at least} one of the resulting copies.

\subsubsection{Conditions on Replication Operation}
\label{sec:rep_con}

In addition to the {\bf Coverage} constraint, we require the replication operation to satisfy the following {\bf Conditions}:
\begin{itemize}
    \item[\bf a)] {\em Joint forwarding and replication:} Replication is performed only on packets to be transmitted.
    \item[\bf b)] {\em Efficient replication:} The destination sets of the created copies do not to overlap.
    \item[\bf c)] {\em Duplication:} Only two copies are created by one replication operation.
    \item[\bf d)] {\em Non-consolidation:} Co-located packets of identical content are {\em not} combined; they are kept as separate copies.
\end{itemize}

It can be shown that these {\bf Conditions} do not reduce the achievable throughput nor increase the minimum attainable cost. 
{\bf Condition a)} avoids replicating packets that are not scheduled for transmission, which only increases network congestion and should be avoided.
{\bf Condition b)} is motivated by the fact that ``receiving multiple packets of identical content at the same destination does not increase network throughput'', and thus replication should be performed in an efficient manner, i.e., NOT producing copies with overlapping destinations, to alleviate network traffic and associated resource consumption.
{\bf Conditions c)} and {\bf d)} are justified in Appendix \ref{apdx:assumptions}.

Combining the {\bf Coverage} constraint and the above {\bf Conditions} leads to the following important property: each destination node of a packet undergoing duplication (we use ``duplication'' instead of ``replication'' in the rest of the paper, e.g., duplication status) must be present in the destination set of exactly one of the two resulting copies.

As illustrated in Fig. \ref{fig:queuing_dynamic}, the duplication operation process works as follows. 
Let $q$ denote the duplication status of a packet selected for (transmission) operation. Upon duplication, one copy is transmitted to the corresponding neighbor node (referred to as the {\em transmitted} copy, of status $s$), and the other copy stays at the node waiting for future operation (referred to as the {\em reloaded} copy, of status $r$). Then, $q = s + r$, and we refer to the pair $(q, s)$ as the {\em duplication choice} (and $r = q - s$). Let $\Omega = \big\{ (q, s): q\in \{0, 1\}^D, s\in 2^q \big\}$ denote the set of all duplication choices.



\begin{remark}
We note that the duplication choice $(q, q)$ describes the special case that a packet is transmitted {\em without duplication}. In particular, note that for a status $b_k$ packet, i.e., a (unicast) packet with one destination, $(b_k, b_k)$ is the {\em only} duplication choice.
\end{remark}

\begin{remark}
\label{remark:des_duplication}
Duplication $(q, b_k)$ automatically takes place when a status $q$ packet with $d_k$ in its current destination set (i.e., $q_k = 1$) arrives at destination $d_k$, in which case: the status $b_k$ copy departs the network immediately, and the status $q - b_k$ copy stays at node $d_k$.
\end{remark}

\vspace{-10pt}
\section{Policy Spaces and Network Stability Region} \label{sec:policy_space}

In this section, we introduce the policy space for multicast packet delivery, based on which we characterize the multicast network stability region.

\vspace{-10pt}
\subsection{Policy Space}

\subsubsection{Decision Variables}

We consider a general policy space for multicast packet delivery, encompassing joint packet forwarding and duplication operations, whose associated forwarding and duplication scheduling variables are described by
\begin{align}
\label{eq:decision_variable}
    \V{\mu}(t) = \big\{ \mu_{ij}^{(q, s)}(t) : (q, s) \in \Omega,\,(i,j) \in \Set{E} \big\}
\end{align}
where $\mu_{ij}^{(q, s)}(t)$ denotes the amount of status $q$ packets selected for (forwarding and duplication) operation, with duplication choice $(q, s)$ and forwarding choice $(i,j)$, at time $t$. That is, $\mu_{ij}^{(q, s)}(t)$ status $q$ packets are duplicated, resulting in $\mu_{ij}^{(q, s)}(t)$ status $s$ packets transmitted over link $(i,j)$, and $\mu_{ij}^{(q, s)}(t)$ status $q-s$ packets reloaded to node $i$, as illustrated in Fig. \ref{fig:queuing_dynamic}.

\subsubsection{Admissible Policies}
\label{sec:admissible_policy}

A control policy is {\em admissible} if the flow variables satisfy:
\begin{enumerate}
    
	\item[a)] non-negativity: $\V{\mu}(t) \succeq 0$, i.e., $\mu_{ij}^{(q, s)}(t) \geq 0$ for $\forall\, (i, j)\in \Set{E}, (q, s)\in \Omega$.
	
	\item[b)] link capacity constraint: $\sum_{(q, s) \, \in \, \Omega} \mu_{ij}^{(q, s)}(t) \leq C_{ij}$ for $\forall\,(i,j)\in \Set{E}$.

	\item[c)] generalized flow conservation: for each intermediate node $i\in \Set{V} \setminus \Set{D}$ and $q\in \{0, 1\}^D$,
	\begin{align} \label{eq:flow_conservation}
	\sum_{s\in 2^{\bar{q}}} \sum_{j\in\delta_i^-} \overline{ \big\{ \mu_{ji}^{(q+s, q)}(t) \big\} } + \sum_{s\in 2^{\bar{q}}} \sum_{j\in\delta_i^+} \overline{ \big\{ \mu_{ij}^{(q+s, s)}(t) \big\} } + \lambda_i^{(q)} = \sum_{s\in 2^q} \sum_{j\in\delta_i^+} \overline{ \big\{ \mu_{ij}^{(q, s)}(t) \big\} }
	\end{align}
	where $\lambda_i^{(q)}$ is the mean rate of exogenously arriving packets $a_i^{(q)}(t) = a_i(t) \, \mathbb{I}\{q = \V{1}\}$,
	and $\bar{q} = \V{1} - q$ denotes the complement of $q$, i.e., the set of destinations not included in $q$.

	\item[d)] boundary conditions: $\mu_{d_k j}^{(q, s)}(t) = 0$ for $\forall\,d_k\in \Set{D},\, j\in \delta^+_{d_k}$, $q$ with $q_k = 1$, $k \in \{1, \cdots, D\}$.
  
\end{enumerate}

\begin{figure}[t]
    \centering
    \includegraphics[width = .95 \columnwidth]{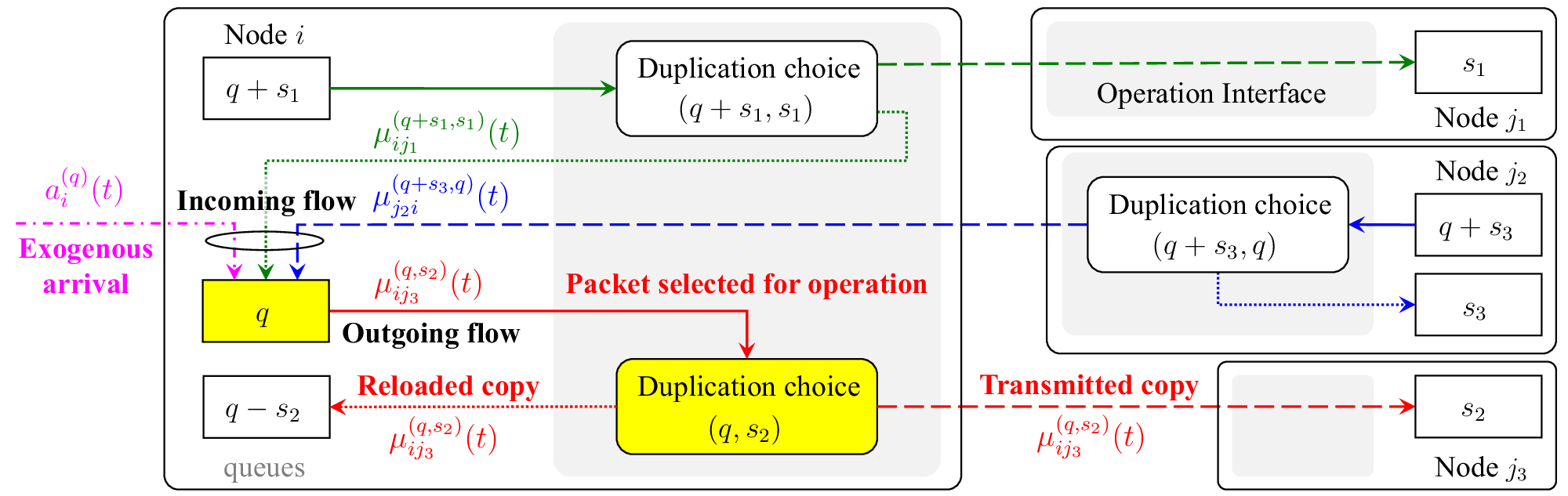}
    \vspace{-10pt}
    \caption{Illustration of joint packet forwarding and duplication operations (solid, dashed, and dotted lines represent packets selected for operation, transmitted copies, and reloaded copies, respectively) and incoming/outgoing flow variables associated with the status $q$ queue of node $i$ (while only explicitly indicated for the red flow, note that all arrows with the same color are associated with the same flow variable).}
    \vspace{-10pt}
    \label{fig:queuing_dynamic}
\end{figure}

The generalized flow conservation c) can be described as follows.
(i) In contrast to the {\em instantaneous} constraints, a), b), and d), which must hold at each time slot, c) imposes an equality constraint on the {\em average flow rates} of incoming/outgoing status $q$ packets to/from node $i$.
(ii) As illustrated in Fig. \ref{fig:queuing_dynamic}, for each node $i$ and status $q$ queue: the incoming flow of status $q$ packets has three components: packets received from each neighbor node $j \in \delta_i^-$ after undergoing duplication $(q+s, q)$ (which creates transmitted copies of status $q$), i.e., $\mu_{ji}^{(q+s, q)}(t)$; packets that stay at node $i$ after undergoing local duplication $(q+s, s)$ (which creates reloaded copies of status $q = (q+s)-s$), i.e., $\mu_{ij}^{(q+s, s)}(t)$; and exogenously arriving status $q$ packets, i.e., $a_i^{(q)}(t)$. The outgoing flow includes all status $q$ packets selected for operation with duplication choice $(q, s)$, i.e., $\mu_{ij}^{(q, s)}(t)$.

\subsubsection{Cost Performance}

The instantaneous overall resource operational cost of an admissible policy is given by 
\begin{align}
\label{eq:cost_model}
    h(t) = h(\V{\mu}(t)) = \sum_{(i, j) \, \in \, \Set{E}} e_{ij} \sum_{(q, s) \, \in \, \Omega} \mu_{ij}^{(q, s)}(t) = \langle \V{e}, \V{\mu}(t) \rangle.
\end{align}
We employ the {\em expected time average cost} $\avg{ \E{h(t)} }$ to characterize the cost performance, in which the expectation is taken \ac{wrt} the flow variables $\V{\mu}(t)$, governed by the arrival process $\V{a}(t)$ and the (possibly randomized) control policy.
We denote by $h^\star(\V{\lambda})$ the minimum attainable cost under the arrival vector $\V{\lambda}$.

\subsection{Multicast Network Stability Region} \label{sec:stability_region}

In this section, we characterize the {\bf multicast network stability region} $\Lambda$ to measure the throughput performance of the system, defined as the set of arrival vectors $\V{\lambda} = \{ \lambda_i : i\in \Set{V} \}$ under which there exists an admissible policy satisfying the constraints in Section \ref{sec:admissible_policy}.

\begin{theorem} \label{thm:stability_1}
An arrival vector $\V{\lambda}$ is interior to the stability region $\Lambda$ if and only if there exist flow variables $\V{f} = \big\{ f_{ij}^{(q, s)} \geq 0 : \forall\, (q, s), \, (i, j) \big\}$ and probability values $\V{\beta} = \big\{ \beta_{ij}^{(q, s)} \geq 0 : \sum_{(q', s')\in \Omega}\beta_{ij}^{(q', s')} \leq 1,\ \forall\, (q, s), \, (i, j)  \big\}$ such that:
\begin{subequations} \label{eq:characterization_1}
\begin{gather}
    \sum_{s\in 2^{\bar{q}}} \sum_{j\in\delta_i^-} f_{ji}^{(q+s, q)} + \sum_{s\in 2^{\bar{q}}} \sum_{j\in\delta_i^+} f_{ij}^{(q+s, s)} + \lambda_i^{(q)} \leq \sum_{s\in 2^q} \sum_{j\in\delta_i^+} f_{ij}^{(q, s)},\  \forall\, i\in \Set{V}, q\in \{0, 1\}^D, \label{eq:c1_1} \\
    f_{ij}^{(q, s)} \leq \beta_{ij}^{(q, s)} C_{ij}, \ \forall\, (i, j)\in \Set{E}, (q, s) \in \Omega, \label{eq:c1_2} \\
    f_{d_k j}^{(q, s)} = 0,\ \forall\,d_k\in \Set{D},\, j\in \delta^+_{d_k},\, q: q_k = 1,\, k \in \{1, \cdots, D\}.
\end{gather}
\end{subequations}
In addition, there exists a stationary randomized policy specified by probability values $\V{\beta}$ (i.e., {\em at each time slot $t$, each link $(i, j)$ selects $C_{ij}$ status $q$ packets for forwarding and duplication operation with duplication choice $(q, s)$ with probability $\beta_{ij}^{(q, s)}$}) and attains optimal cost $h^\star(\V{\lambda})$.
\end{theorem}

\begin{IEEEproof}
The proof follows the same steps as \cite[Appendix A]{FenLloTulMol:J18a} with the following changes: (i) to show {\em Necessity}, we consider a {\em multicast 
stabilizing policy} that makes additional decisions on duplication choices $(q, s)$, 
and (ii) to show {\em Sufficiency}, we consider the {\em stationary randomized policy} presented in Theorem \ref{thm:stability_1}.
The detailed proofs are presented in Appendix \ref{apdx:stability_region}.
\end{IEEEproof}

Next, we analyze the benefit of the multicast framework exploiting in-network packet duplication to {\em enlarge} network stability region, compared to its unicast counterpart $\Lambda_0$.

\begin{proposition} \label{prop:enlarged}
The multicast network stability region $\Lambda$ satisfies $\Lambda_0 \subseteq \Lambda \subseteq D \Lambda_0$, in which $D \Lambda_0 \triangleq \{ D \V{\lambda}: \V{\lambda} \in \Lambda_0 \}$. Furthermore, for any arrival vector $\V{\lambda} \in \Lambda_0$, the minimum attainable cost under the unicast framework, denoted by $h_0^\star(\V{\lambda})$, satisfies $h_0^\star(\V{\lambda}) \geq h^\star(\V{\lambda})$.
\end{proposition}

\begin{IEEEproof}
Since unicast policies, which perform all duplication operations at the source node, become special solutions under the multicast framework, the attained throughput and cost performance cannot be superior to that of the optimal multicast policy, which can eliminate redundant transmissions leveraging in-network packet duplication. Hence, $\Lambda_0 \subseteq \Lambda$ and $h_0^\star(\V{\lambda}) \geq h^\star(\V{\lambda})$. Next, we show $\Lambda \subseteq D \Lambda_0$. For each packet delivered along a multicast route (under a given multicast policy $\mathscr{P}_{\text{m}}$), the unicast policy $\mathscr{P}_{\text{u}}$ can send each copy along the same path as in the multicast route. For each link, $\mathscr{P}_{\text{u}}$ consumes {\em at most $D$ times} the communication resources consumed by $\mathscr{P}_{\text{m}}$ ({\em when $\mathscr{P}_{\text{u}}$ sends $D$ individual copies over a link common to all unicast paths, while $\mathscr{P}_{\text{m}}$ only sends $1$ packet and performs duplication later}). With this key observation, we can show (by contradiction) that the gain factor is bounded by $D$. The detailed proofs are given in 
Appendix \ref{apdx:stability_region}.
\end{IEEEproof}

\section{Problem Formulation}
\label{sec:queuing_system}

\subsection{Queuing System}

Since forwarding and duplication decisions are driven by each packet's destination set, keeping track of data packets' destination sets is essential. A key step is to construct a queuing system with distinct queues for packets of different duplication status $q$. In particular, we denote by $Q_i^{(q)}(t)$ the backlog of the queue holding status $q$ packets at node $i$ at time $t$. Define $Q_i^{(\V{0})}(t) = 0$ and $\V{Q}(t) = \big\{ Q_i^{(q)}(t): i\in \Set{V},\, q \in \{0, 1\}^{D} \big\}$.

Each time slot is divided into two phases. In the {\em transmitting} phase, each node makes and executes forwarding and duplication decisions based on the observed queuing states. In the {\em receiving} phase, the incoming packets, including those received from neighbor nodes, reloaded copies generated by duplication, as well as exogenously arriving packets, are loaded into the queuing system, and the queuing states are updated.

\subsubsection{Queuing Dynamics}

The queuing dynamics are derived for two classes of network nodes.

(i) For an {\em intermediate node} $i\in \Set{V} \setminus \Set{D}$, the queuing dynamics are given by
\begin{align}
\label{eq:q_dynamic}
    Q_i^{(q)}(t+1) \leq \max\Big[ 0, Q_i^{(q)}(t) - \mu^{(q)}_{i\to}(t) \Big] + \mu^{(q)}_{\to i}(t) + a_i^{(q)}(t),\ \forall\, q \in \{0, 1\}^D
\end{align}
where $\mu^{(q)}_{i\to }(t)$ and $\mu^{(q)}_{\to i}(t)$ are the outgoing and (controllable) incoming network flows of status $q$ packets from/to node $i$, given by (as illustrated in Fig. \ref{fig:queuing_dynamic})
\begin{align}
\label{eq:total_flow}
    \mu^{(q)}_{i \to}(t) = \sum_{j\in \delta_i^+} \sum_{s\in 2^q} \mu_{ij}^{(q, s)}(t),\ 
    \mu^{(q)}_{\to i}(t) = \sum_{j\in \delta_i^-} \sum_{s\in 2^{\bar{q}}} \mu_{ji}^{(q+s, q)}(t) + \sum_{j\in \delta_i^+} \sum_{s\in 2^{\bar{q}}} \mu_{ij}^{(q+s, s)}(t).
\end{align}
Note that \eqref{eq:q_dynamic} holds with inequality when the number of status $q$ packets incoming to node $i$ is less than $\mu^{(q)}_{\to i}(t)$ (see Remark \ref{rmk:assigned_flow}), and the difference is bounded by $\sum_{j \in \delta_i^-} C_{ji} + \sum_{j \in \delta_i^+} C_{ij}$.

(ii) For a {\em destination node} $i = d_k\ (k = 1, \cdots, D)$, the queuing dynamics are given by
\begin{align}
\label{eq:q_dynamic_des}
    Q_{i}^{(q)}(t+1)
    \leq \big( I + \mu^{(q+b_k)}_{\to i}(t) 
    \big) \, \mathbb{I}\{ q_k = 0 \}, \ \forall\, q \in \{0, 1\}^D , 
\end{align}
where $I$ represents the right hand side of \eqref{eq:q_dynamic}. To wit, at destination node $d_k$:
(i) all status $q$ queues with $q_k=1$ are always empty, and 
(ii) all other queues have an additional incoming flow corresponding to the reloaded copies resulting from the automatic duplication of status $q+b_k$ packets arriving at the destination (see also Remark \ref{remark:des_duplication}), i.e., $\mu^{(q+b_k)}_{\to i}(t)$.

\begin{remark}
\label{rmk:assigned_flow}
Similar to many existing control policies, e.g., \cite{FenLloTulMol:J18a,TasEph:J92,Nee:B10,cai2020mec}, we note that flow variable $\V{\mu}(t)$ can lead to a total outgoing flow $\mu_{i\to}^{(q)}(t)$ exceeding the available packets in queue $Q_i^{(q)}(t)$, and we include the ramp function $\max[0, \,\cdot\,]$ in \eqref{eq:q_dynamic} to avoid negative queue length. 
If $\mu^{(q)}_{i\to}(t) > Q_i^{(q)}(t)$, the residual capability $\mu^{(q)}_{i\to}(t) - Q_i^{(q)}(t)$ is either used with {\em idle fill} (referred to as dummy packets) or {\em wasted}, which is shown not to
affect the network stability region \cite{Nee:B10}.

\end{remark}

\subsection{Problem Formulation}

The goal is to develop an admissible control policy that stabilizes the queuing system, while minimizing overall operational cost. Formally, we aim to find a control policy with decisions $\{\V{\mu}(t) : t\geq 0\}$ satisfying
\begin{subequations}
\label{eq:formulation}
\begin{align}
    & \min_{\V{\mu}(t) \succeq 0} \quad \avg{ \E{h(t)} } \label{eq:obj_func} \\
    & \hspace{5pt} \operatorname{s.t.} \hspace{10pt} \ \avg{ \E{ \| \V{Q}(t) \|_{1} } } < \infty\text{, i.e., stabilizing } \V{Q}(t) \text{ \eqref{eq:q_dynamic} -- \eqref{eq:q_dynamic_des}}, \label{eq:stability} \\
    & \hspace{40pt} \sum_{(q, s) \, \in \, \Omega} \mu_{ij}^{(q, s)}(t) \leq C_{ij},\ \forall\, (i, j), \label{eq:cap_con_2} \\
    & \hspace{40pt} \mu_{d_k j}^{(q, s)}(t) = 0,\ \forall\,d_k\in \Set{D},\, j\in \delta^+_{d_k},\, q: q_k = 1,\, k \in \{1, \cdots, D\}.
\end{align}
\end{subequations}
In addition, note that by Little's Theorem \cite{Nee:B10}, the average delay is linear in the queue backlog $\avg{ \E{ \| \V{Q}(t) \|_{1} } }$, and thus \eqref{eq:stability} is equivalent to guaranteeing {\em finite average delay}.

\section{Generalized Distributed Cloud Network Control Algorithm}
\label{sec:exact_solution}

In this section, we leverage the \ac{ldp} theory \cite{Nee:B10} to address \eqref{eq:formulation}, which guides the design of the proposed GDCNC algorithm.

\subsection{Lyapunov Drift-plus-Penalty} \label{sec:ldp}

Consider the Lyapunov function $L(t) = \|\V{Q}(t)\|_2^2/2$ and associated Lyapunov drift $\Delta(\V{Q}(t)) = L(t+1) - L(t)$. The LDP approach aims to minimize an upper bound of a linear combination of the Lyapunov drift (whose derivation follows the steps in \cite[Section 3.1.2]{Nee:B10}, and details are shown in 
Appendix \ref{apdx:ldp})
and the objective function weighted by a tunable parameter $V$, i.e.,
\begin{align}
\label{eq:ldp_bound}
    \Delta(\V{Q}(t)) + V h(t)
    \leq |\Set{V}| B + \langle \V{a}(t), \V{Q}(t) \rangle - \langle \V{w}(t), \V{\mu}(t) \rangle
\end{align}
where $B$ is a constant, and the {\em duplication utility weights} $\V{w}(t) = \{ w_{ij}^{(q, s)}(t)\}$ 
are given by
\begin{align}
\label{eq:weight}
    w_{ij}^{(q, s)}(t) = Q_i^{(q)}(t) - Q_i^{(q-s)}(t) - Q_j^{(s)}(t) - V e_{ij}.
\end{align}
Equivalently, the proposed algorithm selects the flow variable $\V{\mu}(t)$ to maximize $\langle \V{w}(t), \V{\mu}(t) \rangle$ at each time slot, which can be decomposed into separate problems for each link $(i, j)$:
\begin{align}
\label{eq:opt_problem}
    \max_{ \mu_{ij}^{(q, s)}(t) \, : \, (q, s) \, \in \, \Omega }\ \sum_{(q, s) \, \in \, \Omega} w_{ij}^{(q, s)}(t) \mu_{ij}^{(q, s)}(t),\ 
    \st \  \sum_{(q, s) \, \in \, \Omega} \mu_{ij}^{(q, s)}(t) \leq C_{ij},\  \mu_{ij}^{(q, s)}(t) \geq 0.
\end{align}
The resulting {\em max-weight} solution is described in the next section.

\subsection{Generalized Distributed Cloud Network Control} \label{sec:control_alg}

\begin{algorithm}[t]
\begin{algorithmic}[1]
    
    \FOR{
    $t \geq 0$ and $(i, j) \in \Set{E}$
    }
    
    \STATE{
    Calculate the duplication utility weight $w_{ij}^{(q, s)}(t)$ for all duplication choices $(q, s)$ 
    by \eqref{eq:weight}.
    }
    
    \STATE{
    Find the $(q, s)$ pair with the largest weight $(q^\star, s^\star) = \argmax_{(q, s) \, \in \, \Omega} \, w_{ij}^{(q, s)}(t)$.
	}
	
    \STATE{
    Assign transmission flow: $\mu_{ij}^{(q, s)}(t) = C_{ij} \, \mathbb{I}\big\{ w_{ij}^{(q^\star, \, s^\star)}(t) > 0, (q, s) = (q^\star, s^\star) \big\}$.
	}
    \ENDFOR
\end{algorithmic}
\caption{GDCNC}
\label{alg:GDCNC}
\end{algorithm}

The developed 
{\em generalized distributed cloud network control} (GDCNC) algorithm is described in Algorithm \ref{alg:GDCNC}, and exhibits two salient features.
(i) {\em Decentralized}: it only requires local information exchange (i.e., queuing states of neighbor nodes) and decision making, which can be implemented in a fully distributed manner.
(ii) {\em Sparse}: for each link $(i, j) $, it selects one duplication choice $(q^\star, s^\star)$ at each time slot, which affects the states of status $q^\star$ and $q^\star - s^\star$ queues at node $i$, and status $s^\star$ queue at node $j$.
Therefore, for each node $i$, the number of queues with changing states in one time slot is $2|\delta_i^+| + |\delta_i^-| \sim \mathcal{O}(|\delta_i|)$, where $|\delta_i|$ denotes the degree of node $i$.

\begin{remark}
Note that when dealing with a unicast flow, where all packets have the same single destination, the only valid duplication status is $q=1$ and the only valid duplication choice is $(q, s) = (1, 1)$, in which case GDCNC reduces to DCNC \cite{FenLloTulMol:J18a}.
\end{remark}

\subsection{Performance Analysis}

In this section, we analyze the delay and cost performance of GDCNC, and its complexity from both communication and computation dimensions.

\subsubsection{Delay-Cost Tradeoff}

In the following theorem, we employ the minimum attainable cost $h^\star(\V{\lambda})$ as the benchmark to evaluate the performance of GDCNC.

\begin{theorem} \label{thm:tradeoff}
For any arrival vector $\V{\lambda}$ interior to the multicast stability region $\Lambda$, the average queue backlog and the operational cost achieved by GDCNC satisfy
\begin{gather}
\label{eq:tradeoff}
    \avg{ \E{\| \V{Q}(t) \|_1} } \leq \frac{|\Set{V}| B}{\epsilon} + \left[ \frac{h^\star(\V{\lambda}+\epsilon\V{1}) - h^\star(\V{\lambda}) }{ \epsilon } \right] V,\quad
    \avg{ \E{h(t)} } \leq h^\star(\V{\lambda}) + \frac{|\Set{V}| B}{V},
\end{gather}
for any $\epsilon > 0$ such that $\V{\lambda}+\epsilon\V{1}\in \Lambda$.
\end{theorem}

\begin{IEEEproof}
Applying {\em expected time average} and {\em telescope sum} \cite{Nee:B10} to \eqref{eq:ldp_bound}, we can obtain $V \avg{ \E{h(t)} } \leq |\Set{V}| B - \epsilon \avg{ \E{ \| \V{Q}(t) \|_1 } } + V h^\star( \V{\lambda} + \epsilon \V{1} )$.
Then, using $\avg{ \E{h(t)} } \geq h^\star(\V{\lambda})$ and $- \epsilon \avg{ \E{ \| \V{Q}(t) \|_1 } } \leq 0$ leads to \eqref{eq:tradeoff}.
See 
Appendix \ref{apdx:tradeoff}
for a detailed derivation.
\end{IEEEproof}

The above theorem is illustrated as follows. For any arrival vector interior to the stability region, GDCNC (using any fixed $V \geq 0$) can stabilize the queuing system, and thus is throughput-optimal. In addition, GDCNC achieves an $[\mathcal{O}(V), \mathcal{O}(1/V)]$ tradeoff between delay (which is linear in queue backlog) and cost: by pushing $V \to \infty$, the attained cost can be arbitrarily close to the minimum $h^\star(\V{\lambda})$, with a tradeoff in network delay.

\subsubsection{Complexity Issues}
\label{sec:complexity}

Next, we analyze the complexity of GDCNC.

{\bf Communication Overhead:}
At each time slot, GDCNC requires local exchange of queue backlog information. 
Instead of the entire queuing states (of size $| \{0, 1\}^D | \sim \mathcal{O}(2^D)$), we can leverage {\em sparsity} (see Section \ref{sec:control_alg}) to reduce communication overhead, i.e., nodes only exchange information of the queues with changing states, reducing the overhead to $\mathcal{O}(|\delta_i|)$.

{\bf Computational Complexity:}
At each time slot, GDCNC calculates the utility weight of each duplication choice $(q, s)$, of computational complexity proportional to $|\Omega| \sim \mathcal{O}(3^D)$ (see
Appendix \ref{apdx:complexity}),
i.e., exponential in the destination set size.
This is due to the combinatorial nature of the multicast routing problem. Indeed, the state-of-the-art centralized solution to the multicast flow control problem \cite{zhang2021multicast} requires solving the NP-complete Steiner tree problem for route selection at each time slot.

\section{GDCNC-R with Reduced Complexity} \label{sec:apx_solution}

In this section, we develop GDCNC-R, a variant of GDCNC achieving sub-optimal performance with {\em reduced} complexity.

\vspace{-10pt}
\subsection{Duplication Tree} \label{sec:dup_tree}

\begin{figure}[t]
    \centering
    \includegraphics[width = .95 \columnwidth]{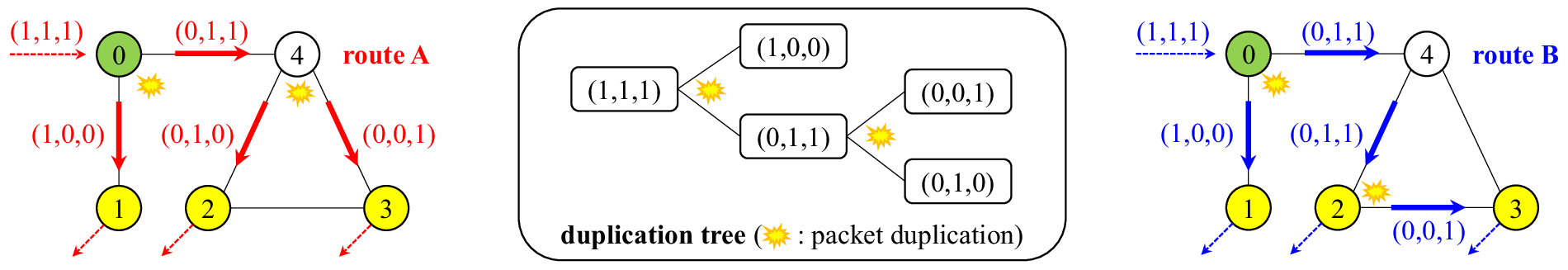}
    \vspace{-10pt}
    \caption{
        Different multicast routes (i.e., routing trees) can share the same duplication choices (i.e., duplication tree). 
        For example, the red and blue routes used to deliver the status $(1,1,1)$ packet from source node $0$ to destination nodes $\{ 1, 2, 3 \}$ are associated with the same duplication tree.
    }
    \vspace{-25pt}
    \label{fig:duplication_tree}
\end{figure}

To illustrate the complexity reduction of GDCNC-R, we first define the duplication tree as a useful representation of the duplication operations performed on a given multicast packet.

\begin{definition}
A {\em duplication tree} $\mathcal{T}$ is a binary tree with each node associated with a duplication status:
the root node and the $D$ leaf nodes represent the status of the initial multicast packet, $\V{1}$, and of the copies delivered to each destination, $b_k\ (k \in \{1, \cdots, D\})$, respectively. Each internal node $q$ splits into two child nodes $s$ and $r$, associated with duplication $(q, s)$ and $(q, r)$, respectively.
\end{definition}

As illustrated in Fig. \ref{fig:duplication_tree}, the duplication choices of a given multicast route are described by a duplication tree. Note however that different routes can share the same duplication choices, hence the same duplication tree. GDCNC achieves optimal performance by evaluating all possible duplication choices at each node, which is equivalent to consider all routes in all duplication trees. This is also the reason for its high computational complexity.

\vspace{-10pt}
\subsection{Proposed Approach}

In contrast, the developed GDCNC-R algorithm narrows the focus on the multicast routes included in a subset of {\em effective} duplication trees and associated duplication choices.
Given a duplication tree, the policy can select from all the multicast routes mapping into the tree for packet delivery, with the goal of reducing network congestion, and as a result, improving the delay performance.
The selected duplication trees shall minimize the resulting {\em (throughput and cost) performance loss}, and the proposed approach is referred to as {\em destination clustering}.

Importantly, the main benefit of the multicast framework is to save redundant operations by exploiting in-network packet duplication. To maximize this gain, a duplication shall be performed only on packets with ``distant'' destinations in their current destination set (under the distance metrics listed 
at the end of this section).

To this end, we propose to construct the duplication tree as follows
(in this paragraph, we refer to a node in the duplication tree as a {\em vertex} to distinguish from a node in the routing network).
We start from the entire network node set $\Set{V}$, which includes all destinations and corresponds to the root vertex of the duplication tree. At each step, the current node set is divided into {\em two} disjoint clusters -- in line with the {\em duplication} condition (see Section \ref{sec:rep_con}). The two child vertices are given by the destinations in the resulting clusters. Then, we move to the child vertices and repeat this step for the corresponding clusters, until each cluster only includes one destination node (we discard clusters not including any destinations in the procedure).

While there are many existing clustering methods, we use the widely adopted {\em $k$-means clustering} \cite{alpaydin2020ml},
described as follows.
Given a set of data points with known {\em features} and a distance function measuring the {\em similarity} between two data points, the algorithm starts assigning random labels (i.e., the cluster to which the data point belongs) to the data points; then, it (i) calculates the {\em center} for each cluster, and (ii) updates the label of each data point to the cluster with the {\em nearest} center (under the given distance function) in an iterative manner until converging.

Some effective choices for the inter-node distance are listed as follows. In the wired case: To optimize the throughput performance, we can use the {\em reciprocal transmission capacity} as the inter-node distance; while {\em unit transmission cost} is a proper metric to deal with the cost performance. 
For nodes that are not directly connected, we assume that there is a link of zero capacity and infinity cost between them.
Given the pairwise similarities, we can apply spectrum clustering techniques \cite{Meila2007clustering} to define features for the data points, followed by $k$-means clustering.
In the wireless scenario, a straightforward distance metric is the {\em Euclidean distance}, which leads to using the {\em geographical location} of each network node as its feature, and whose effectiveness is validated by numerical experiments (see Fig. \ref{fig:stability_wo_pa} and \ref{fig:stability_w_pa}).

\subsection{Complexity Analysis}

As shown in
Appendix \ref{apdx:complexity},
each duplication tree includes $2D - 1$ nodes, with $D-1$ internal nodes (including the root node) and $D$ leaf nodes. There are $3$ possible duplication choices associated with each internal node $q$ (with child nodes $s$ and $r$), i.e., $(q, q)$, $(q, s)$ and $(q, r)$, while only one duplication choice $(b_k, b_k)$ is associated with each leaf node $b_k$. Therefore, every duplication tree includes $4D - 3 \sim \mathcal{O}(D)$ possible duplication choices.

GDCNC-R uses $K$ duplication trees, with $K$ chosen to strike a good balance between performance optimality and computational complexity. Specifically, GDCNC-R has $\mathcal{O}(KD)$ complexity, i.e., polynomial in the number of duplication trees and destination set size.

\section{Extensions}
\label{sec:ext}

In this section, we present extensions to the GDCNC algorithms, including: (i) a cloud network control policy for multicast \ac{agi} service delivery, (ii) a modified algorithm for \ac{mec} scenarios (with wireless links), and (iii) a variant of GDCNC with {\em Enhanced} delay performance, EGDCNC.

\subsection{Multicast AgI Service Delivery}
\label{sec:AgI}

In line with \cite{FenLloTulMol:J18a,zhang2021multicast}, the additional packet processing decisions involved in the distributed cloud network setting can be handled as follows.

\subsubsection{Cloud Network Model}

Consider a cloud network composed of compute-enabled nodes that can process data packets via corresponding service functions, with the available processing resources and associated costs defined as: (i) {\em processing capacity} $C_i$, i.e., the computation resource units (e.g., computing cycles per time slot) at node $i$, and (ii) {\em unit processing cost} $e_i$, i.e., the cost to run one unit of computation resource in one time slot at node $i$.

\subsubsection{AgI Service Model}

Consider an \ac{agi} service modeled as \ac{sfc}, i.e., a chain of $M - 1$ functions, through which source packets must be processed to produce consumable results, resulting in the end-to-end data stream divided into $M$ {\em stages}. Each processing step can take place at different network locations hosting the required service functions, and each function $(m\in \{1, \cdots, M-1\})$ in the service chain is described by two parameters: (i) {\em scaling factor} $\xi^{(m)}$, i.e., the number of output packets per input packet, and (ii) {\em workload} $r^{(m)}$, i.e., the amount of computation resource to process one input packet.

\subsubsection{Processing Decision}

We create different queues to hold packets of different processing stages and duplication status. Let $Q_i^{(m, q)}(t)$ denote the backlog of (stage $m$, status $q$) packets, and $\mu_i^{(m, q, s)}(t)$ the scheduled {\em processing flow}, i.e., the amount of (stage $m$, status $q$) packets selected for (processing and duplication) operation with duplication choice $(q, s)$ at node $i$ at time $t$.

Following the procedure in Section \ref{sec:ldp} on a properly constructed cloud layered graph \cite{zhang2021multicast} (see
Appendix \ref{apdx:AgI}
for the full derivation),
the processing decisions for each node $i$ at time $t$ are given by
\begin{enumerate}
	\item[(i)] Calculate the duplication utility weight $w_i^{(m, q, s)}(t)$ for each $(m, q, s)$ tuple:
	\begin{align}
	w_i^{(m, q, s)}(t) = \big[ Q_i^{(m, q)}(t) - \xi^{(m)} Q_i^{(m+1, s)}(t) - Q_i^{(m, q-s)}(t) \big] \big/ r^{(m)} - V e_i \, .
	\end{align}
	\item[(ii)] Find the $(m, q, s)$ tuple with the largest weight: $(m^\star, q^\star, s^\star) = \argmax_{(m, q, s)} \, w_i^{(m, q, s)}(t)$.
	\item[(iii)] Processing flow $\mu_i^{(m,\, q,\, s)}(t) = 
	\big( C_i / r^{(m)} \big) \, \mathbb{I}\big\{ w_i^{(m^\star,\, q^\star,\, s^\star)}(t) > 0, (m, q, s) = (m^\star, q^\star, s^\star) \big\}.$
\end{enumerate}

\subsection{MEC Scenario}
\label{sec:mec}

In line with \cite{cai2020mec}, the packet transmission decisions can be modified to handle wireless distributed cloud network settings, e.g., MEC \cite{CheHao:J18}, as follows.

\subsubsection{Wireless Transmission Model}

Consider a MEC network composed of two types of nodes, i.e., \acp{ue} and \acp{es}, collected in $\Set{V}_{\text{UE}}$ and $\Set{V}_{\text{ES}}$, respectively. Each \ac{es}, equipped with massive antennas, is assigned a separate frequency band of width $B_0$, and with the aid of beamforming techniques, it can transmit/receive data to/from multiple \acp{ue} simultaneously without interference (assuming that the \acp{ue} are spatially well separated) \cite{AdhNamAhnCai:J13}; while each \ac{ue} is assumed to associate with only one \ac{es} at each time slot. For each wireless link $(i, j)$, the transmission power $p_{ij}(t)$ is assumed to be constant during a time slot (and the maximum power budget of node $i$ is denoted by $P_i$), incurring cost $(\tilde{e}_i \tau) \, p_{ij}(t)$, with $\tilde{e}_i$ denoting the {\em unit energy cost} at node $i$ and $\tau$ the time slot length. Besides, we assume that the channel gain $g_{ij}(t)$ is i.i.d. over time and known by estimation, and the noise power is denoted by $\sigma_{ij}^2$\,.
The \acp{es} are connected by wired links, as described in Section \ref{sec:routing_net}.

\subsubsection{Wireless Transmission Decision}

In addition to flow assignment $\V{\mu}(t) = \{ \mu_{ij}^{(q, s)}(t) \geq 0: (i, j), (q, s)\}$, the control policy makes decisions on power allocation, i.e., $\V{p}(t) = \{ p_{ij}(t) \geq 0: (i, j) \}$,
and \ac{ue}-\ac{es} association, 
i.e., $\V{\chi}(t) = \{ \chi_{ij}(t) \in \{0, 1\} : (i, j) \}$, where $\chi_{ij}(t)$ indicates if 
node $i$ is associated with node $j$ ($\chi_{ij}(t) = 1$) or not ($\chi_{ij}(t) = 0$).

The LDP bound \eqref{eq:ldp_bound} remains valid, and the resulting problem \eqref{eq:opt_problem} is given by
\begin{subequations}
\label{eq:mec}
\begin{align}
    & \max \ \sum\nolimits_{(i, j)\in \Set{E}} \chi_{ij}(t) \Psi_{ij}(t),\ \Psi_{ij}(t) \triangleq \sum\nolimits_{(q, s)\in \Omega} w_{ij}^{(q, s)}(t) \mu_{ij}^{(q, s)}(t) - V (\tilde{e}_i \tau) p_{ij}(t) \\
    & \, \st \hspace{5pt} \sum\nolimits_{j\in \delta_i^+} \chi_{ij}(t) \leq 1\text{ for }\forall\, i\in \Set{V}_{\text{UE}};\ 
    \chi_{ij}(t) = 1 \text{ for } \forall\, i\in \Set{V}_{\text{ES}}, \, j \in \delta_i^+, \label{eq:association} \\
    & \hspace{28pt} \sum\nolimits_{(q, s)\in \Omega} \mu_{ij}^{(q, s)}(t) \leq \tau R_{ij}(t), 
    \ R_{ij}(t) \triangleq B_0 \log_2\big( 1 + g_{ij}(t) \, p_{ij}(t) \big/ \sigma_{ij}^2 \big),\ \forall\, (i, j) \\ 
    & \hspace{28pt} \sum\nolimits_{j\in \delta_i^+} \chi_{ij}(t) \, p_{ij}(t) \leq P_i,\ \forall\, i \in \Set{V} 
\end{align}
\end{subequations}
where $w_{ij}^{(q, s)}(t) = Q_i^{(q)}(t) - Q_i^{(q-s)}(t) - Q_j^{(s)}(t)$,
and \eqref{eq:association} represents the constraint that each \ac{ue} can only select one \ac{es} to communicate with at each time slot.
Following the procedure in \cite{cai2020mec}, we can solve \eqref{eq:mec} to derive the wireless transmission decisions for each link $(i, j)$ at time $t$ as
\begin{subequations}
\label{eq:mec_solution}
\begin{align}
    & \text{Power allocation: }
    p_{ij}^\star(t) = \min\big[ p_{ij}(t; 0), P_i\big],\ i\in \Set{V}_{\text{UE}};\ 
    p_{ij}^\star(t) = p_{ij}(t; \nu^\star),\ i\in \Set{V}_{\text{ES}}, \\
    & \text{Flow assignment: }
    \mu_{ij}^{\star \, (q, s)}(t) = \tau R_{ij}(t) \, \mathbb{I}\{ (q, s) = (q^\star, s^\star) \}, \ (q^\star, s^\star) = \argmax_{(q, s)} \, w_{ij}^{(q, s)}(t), \\ 
    & 
    \text{\ac{ue}-\ac{es} association: }
    \chi_{ij}^\star(t) = \mathbb{I}\{ j = j^\star, \Psi_{ij}^\star(t) > 0 \},\ i\in \Set{V}_{\text{UE}};\ 
    \chi_{ij}^\star(t) = 1,\ i\in \Set{V}_{\text{ES}}
\end{align}
\end{subequations}
where $p_{ij}(t; \nu) \triangleq \max \big[ B_0 w_{ij}^{(q^\star,\, s^\star)}(t) \big/ \big( \tilde{e}_i V + \nu \big) \big/ \ln 2 - \sigma_{ij}^2 \big/ g_{ij}(t) , 0 \big]$, $\nu^\star = \max\{0, \nu_0 \}$ with $\nu_0$ satisfying $\sum_{j\in\delta_i^+}$ $p_{ij}(t; \nu_0) = P_i$, and $j^\star = \argmax_{j} \Psi_{ij}^\star(t)$.

\subsection{EGDCNC with Enhanced Delay}

In line with \cite{YinShaRedLiu:J11,FenLloTulMol:J18a}, a {\em biased} queue that incorporates network topology information can be designed to enhance the delay performance of multicast flow control.

In the unicast setting, \cite{YinShaRedLiu:J11} defines the bias term for each node $i$, $H_{\text{U}}(i, d)$, as the {\em minimum hop-distance} between node $i$ and destination $d$, and combines it with the physical queue as $\tilde{Q}_i(t) = Q_i(t) + \eta\, H_{\text{U}}(i, d)$. The bias term creates an intrinsic pressure difference that pushes packets along the shortest path to the destination. The parameter $\eta$ can be found by grid search to optimize the combined effect of hop-distance and queue backlog on the total network delay.

In the multicast setting, we propose to modify the biased queue as
\begin{align} \label{eq:bias}
    \tilde{Q}^{(q)}_i(t) = \|q\|_1 Q_i^{(q)}(t) + \eta\, H_{\text{M}}(i, q),\text{ with }
    H_{\text{M}}(i, q) \triangleq \sum_{k=1}^{D} q_k \, H_{\text{U}}(i, d_k).
\end{align}
Differently from the unicast case, (i) the bias term $H_{\text{M}}(i, q)$ is now the {\em sum of minimum hop-distances} to all current destinations, and (ii) the backlog term $Q_i^{(q)}(t)$ is now weighted by $\|q\|_1$ 
in line with its contribution to the average delay (i.e., the delay a status $q$ packet experiences will impact the delay of the $\|q\|_1$ copies resulting from its subsequent duplications; the detailed proof is given in
Appendix \ref{apdx:delay}.

EGDCNC works just like GDCNC, but using $\tilde{\V{Q}}(t)$ in place of $\V{Q}(t)$ to make forwarding and duplication decisions. Following the procedure in \cite{YinShaRedLiu:J11}, we can show that EGDCNC (using any fixed $\eta \geq 0$) does not lose throughput and cost optimality as compared to GDCNC.

\section{Numerical Results} \label{sec:experiments}
\vspace{-5pt}

\subsection{Network Setup}

We consider a \ac{mec} network within a square area of $200\ \text{m}\times 200\ \text{m}$, including $9$ \acp{ue} and $4$ \acp{es}. In a Cartesian coordinate system with the origin located at the square center, the \acp{es} are placed at $(\pm 50\ \text{m},\pm 50\ \text{m})$. Each user is moving according to Gaussian random walk (reflected when hitting the boundary), and the displacement in each time slot distributes in $\mathcal{N}(\V{0}, 10^{-4} \, \M{I})$~m. The length of each time slot is $\tau = 1\ \text{ms}$.

\begin{table}[t]
    \centering
    \caption{\vspace{-5pt}Network Resources and Operational Costs of the Studied System\vspace{-15pt}}
    
    \begin{tabular}{|l|l|l|l|}
    \hline
    \multicolumn{2}{|c|}{Processing} & \multicolumn{2}{c|}{Transmission} \\ \hline
        \ac{ue} & $C_i = 1$ GHz, $e_i = 2\,/$GHz          &  \ac{ue}-ES         &  $i\in \Set{V}_\text{UE}$: $P_i = 200\,$mW, $\tilde{e}_i = .01 \,/$J; $i\in \Set{V}_\text{ES}$: $P_i = 1\,$W, $\tilde{e}_i = .005\,/$J         \\ \hline
        ES    & $C_i = 5$ GHz, $e_i = 1\ /$GHz          &    ES-ES       &    $C_{ij} = 1$ Gbps, $e_{ij}$ = $1\,/$Gbps for $(i, j)$ if
        $\text{distance}(i, j) = 100\,$m \\
        \hline
    \end{tabular}
    \label{tab:resource}
    \vspace{-25pt}
\end{table}

We employ the transmission model described in Section \ref{sec:mec}, where \acp{es} are connected by wired links, while \acp{ue} and \acp{es} communicate via wireless links with the following parameters: bandwidth $B_0 = 100\ \text{MHz}$, path-loss $= 32.4 + 20 \log_{10}(f_{\text{c}}) + 31.9 \log_{10}(\text{distance})\ \text{dB}$ with $f_{\text{c}} = 30$ GHz (urban microcell \cite{3gpp2018study}), communication range $= 150$~m, standard deviation of shadow fading $\sigma_{\text{SF}} = 8.2\ \text{dB}$, and $\sigma_{ij}^2 = N_0 B_0$ with noise spectral density $N_0 = -174\ \text{dBm}/\text{Hz}$. The processing/transmission capacities and associated costs are shown in Table \ref{tab:resource}.

Consider two AgI services composed of $2$ functions with parameters given by (the subscript of each parameter denotes the associated service $\phi$, and workload $r_{\phi}^{(m)}$ is in GHz$/$Mbps):
\begin{align*}
    \xi_1^{(1)} = 1,\, r_1^{(1)} = \frac{1}{300},\, \xi_1^{(2)} = 2,\, r_1^{(2)} = \frac{1}{400};\quad
    \xi_2^{(1)} = \frac{1}{3},\, r_2^{(1)} = \frac{1}{200},\, \xi_2^{(2)} = \frac{1}{2},\, r_2^{(2)} = \frac{1}{100}.
\end{align*}
Each service has $1$ source node and $D = 3$ destination nodes, which are randomly selected from the \acp{ue}, and the arrival process is modeled as i.i.d. Poisson with $\lambda$ Mbps.

\vspace{-15pt}
\subsection{Uniform Resource Allocation}
\vspace{-5pt}

First, we compare the proposed GDCNC algorithms with two state-of-the-art cloud network control policies: UCNC (a throughput-optimal source routing algorithm) \cite{zhang2021multicast} and EDSPA (a widely used multicast routing technique) \cite{ananta2014multicasting}. Since resource allocation exceeds the scope of the benchmark algorithms, the following policy is employed for all algorithms for fair comparisons: each node/link uses maximum processing/transmission capacity, and each \ac{es} allocates equal transmission power for each \ac{ue}. In the GDCNC algorithms, we select $V = 0$ to optimize the delay performance; in GDCNC-R, we select $K = 1$ duplication tree by $k$-means clustering using geographic distance; in UCNC, route selection is based on delayed information resulting from hop-by-hop transmission of the queuing states from all nodes to the source.

\begin{figure}[t]
    \centering
    \subfloat[Uniform allocation (benchmark: EDSPA, UCNC).]{
    \includegraphics[width = .4 \linewidth]{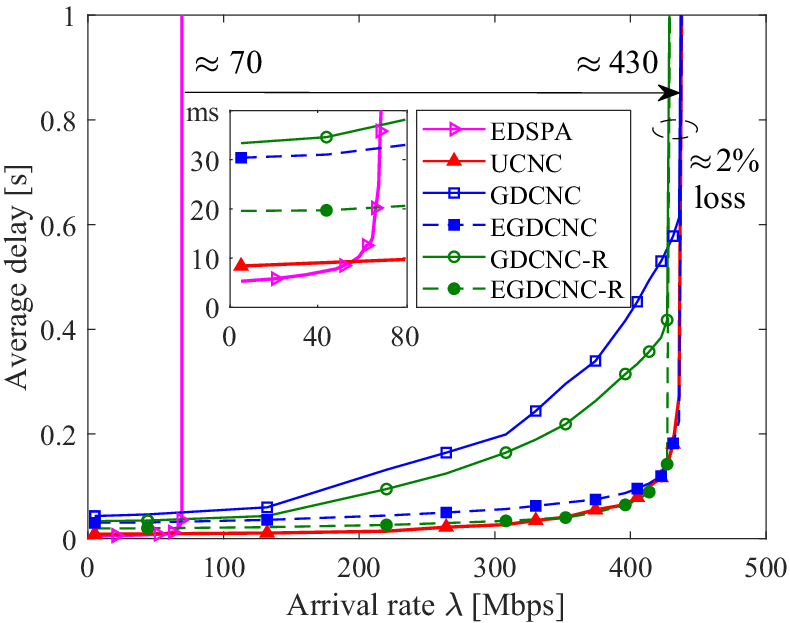}
    \label{fig:stability_wo_pa}
    }
    \hspace{40pt}
    \subfloat[Optimal allocation (benchmark: DCNC, EDCNC).]{
    \includegraphics[width = .4 \linewidth]{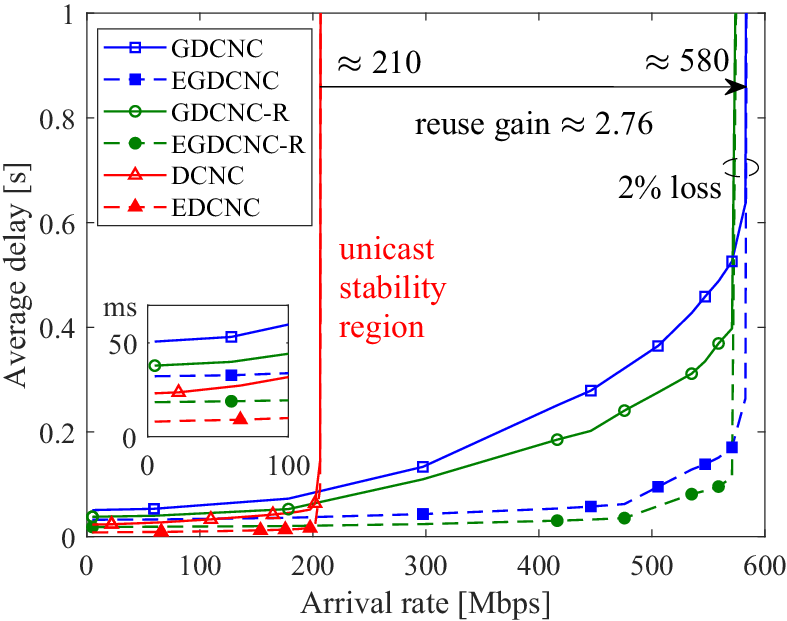}
    \label{fig:stability_w_pa}
    }
    \caption{Network stability regions attained by the benchmark and proposed algorithms.}
    \vspace{-30pt}
\end{figure}

Fig. \ref{fig:stability_wo_pa} depicts the average delay attained by the algorithms under different arrival rates.
First, we focus on the throughput performance. We observe an identical critical point ($\approx 430$ Mbps) for GDCNC, EGDCNC, and UCNC, at which point the average delay blows up, indicative of the stability region boundary, validating Proposition \ref{thm:tradeoff}, i.e., the throughput-optimality of GDCNC/EGDCNC (because UCNC is throughput-optimal \cite{zhang2021multicast}). GDCNC-R/EGDCNC-R achieve sub-optimal throughput by only $\approx 2\%$, illustrating the marginal performance loss from precluded duplication trees. Finally, EDSPA only achieves a maximum rate $\approx 70$ Mbps, which is far from the stability region boundary, due to the lack of route diversity.

When looking at the delay performance, we observe that the enhanced variants EGDCNC/ EGDCNC-R effectively reduce the delay of the initial solutions GDCNC/GDCNC-R, by adding the shortest path bias to improve the delay performance.
In low-congestion regimes, the end-to-end delay mainly depends on the hop-distance of the route; as a result, EDSPA, which delivers the packet along the {\em shortest} path, and UCNC, which requires the entire path to be {\em acyclic} for hop-distance reduction, outperform EGDCNC/EGDCNC-R.
As arrival rate increases, queuing delay becomes the dominant component, and gaps between EGDCNC/EGDCNC-R and UCNC vanish, demonstrating the good performance of the algorithms in high-congestion regimes.

Finally, we study the complexity of the algorithms.
We first sort the algorithms by {\em communication overhead}, i.e., required information for decision making. Noting that network topology changes much slower than queuing states, we have: EDSPA (requiring network topology) $\ll$ GDCNC (requiring local queuing states) $\approx$ EGDCNC (requiring local queuing states and network topology) $\ll$ UCNC (requiring global queuing states).
We then present the running time of the algorithms as a measure for {\em computational complexity}: EDSPA ($1.3$\,s) $<$ GDCNC $\approx$ EGDCNC ($2.7$\,s) $\ll$ UCNC ($5.8$\,min), among which: EDSPA runs fastest because it uses fixed routes; GDCNC can operate efficiently to complete simple algebraic operations \eqref{eq:weight}; and UCNC requires solving the NP-complete Steiner tree problem (in a $39$-node layered graph) at each time slot, which incurs high computational complexity that can increase when applied to larger networks.

To sum up: in low-congestion regimes, EDSPA is a good choice due to its low complexity and superior delay performance; in high-congestion regimes, UCNC works better for small networks that impose low overhead for information collecting and decision making; and EGDCNC is competitive in all regimes and especially suitable for large-scale distributed cloud networks.

\vspace{-10pt}

\subsection{Optimal Resource Allocation}
\label{sec:exp_wp}

Next, we demonstrate the cost performance of the GDCNC algorithms, employing DCNC \cite{FenLloTulMol:J18a} as the benchmark, which is throughput- and cost-optimal for unicast 
flow control.

\subsubsection{Network Stability Region}

Fig. \ref{fig:stability_w_pa} shows the network stability regions attained by the algorithms. We find that: the designed power allocation policy \eqref{eq:mec_solution} boosts the stability region ($\approx 580$ Mbps), compared to {\em uniform} power allocation ($\approx 430$ Mbps, as shown in Fig. \ref{fig:stability_wo_pa}). Another observation is: compared to DCNC (which attains the {\em unicast} stability region $\approx 210$ Mbps), in the studied setting (with $D = 3$ destinations), the multicast framework enlarges the stability region by a factor of $2.76$, which is bounded by 
$D$, validating Proposition \ref{prop:enlarged}. Finally, the performance loss of GDCNC-R is marginal ($\approx 2\%$) compared to GDCNC, which, together with the results shown in Fig. \ref{fig:stability_wo_pa}, validates its competitive throughput performance.

\subsubsection{Delay and Cost Performance}

Next, we present the tunable delay and cost performance of the algorithms, under $\lambda = 150$ Mbps and $\eta = 0$ (we fix $\eta$ and focus on the effects of $V$).

\begin{figure}
\centering
\subfloat[Average delay (versus $V$).]{
    \includegraphics[width = .3 \columnwidth]{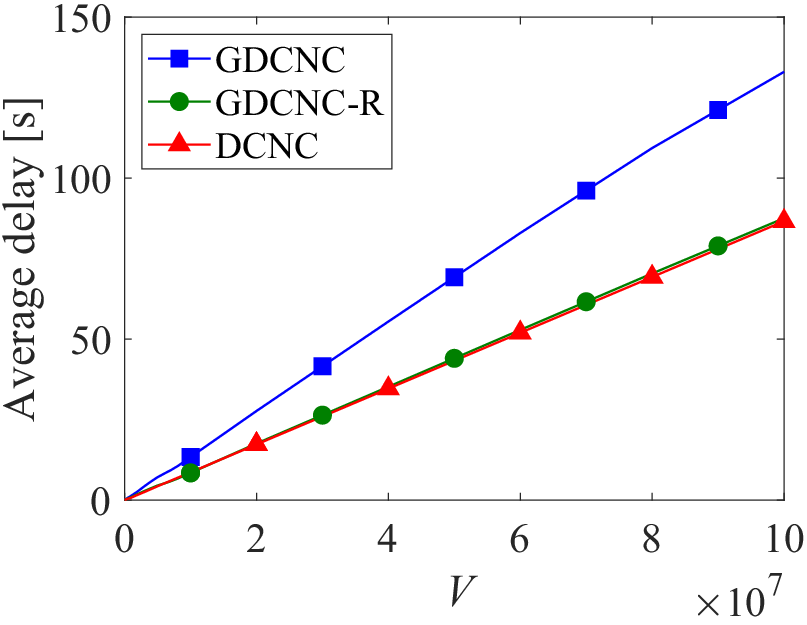}
    \label{fig:delay}
    }
    \hfill
    \subfloat[Operational cost (versus $V$).]{
    \includegraphics[width = .3 \columnwidth]{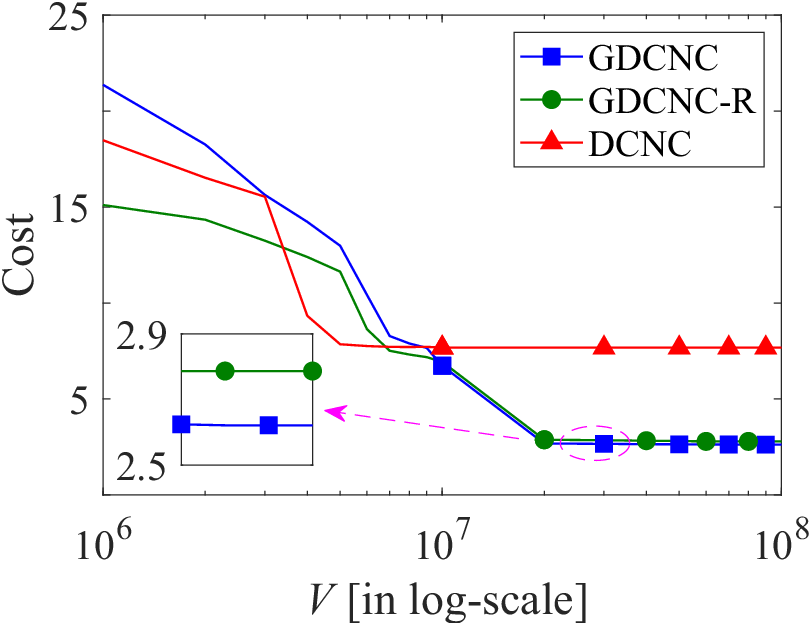}
    \label{fig:cost}
    }
    \hfill
    \subfloat[Effects of destination set size $D$.]{
    \includegraphics[width = 0.3 \columnwidth]{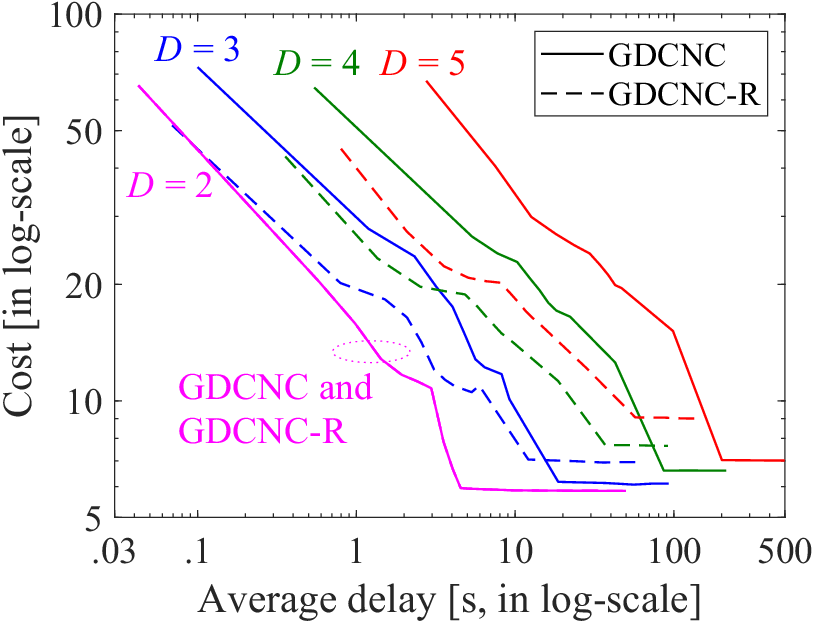}
    \label{fig:tradeoff_2}
    }
    \caption{Delay-cost tradeoffs attained by the three algorithms under different $V$ parameters and destination set sizes $D$.}
    \vspace{-20pt}
\end{figure}

As shown in Fig. \ref{fig:delay}, the average delay attained by each algorithm increases linearly with $V$. Note however that DCNC can achieve a better delay than GDCNC (and almost the same as GDCNC-R) for arrival rates within the unicast stability region (low congestion regime in the enlarged multicast stability region), because it can select separate paths for each copy to optimize the individual delays. This is expected, as the throughput-optimal design of GDCNC, which {\em jointly} selects the copies' paths to reduce network traffic and enlarge the stability region, can lead to higher delay in low-congestion regimes. 

Fig. \ref{fig:cost} shows the reduction in operational cost with growing $V$, validating Theorem \ref{thm:tradeoff}. By pushing $V \to \infty$, the curves converge to the corresponding optimal costs, given by: DCNC ($7.7$) $>$ GDCNC-R ($2.8$) $\approx$ GDCNC ($2.65$). For $V < 10^7$, GDCNC-R attains an even lower cost than GDCNC, even though it has sub-optimal {\em asymptotic} performance (e.g., $V > 10^7$). However, note that a large $V$ also leads to an excessive delay, making it a sub-optimal choice in practical systems. For example, when increasing $V$ from $10^6$ to $10^7$, the cost attained by GDCNC reduces from $21.4$ to $6.7$, while the delay grows from $1.3$ to $13.4$ seconds (these results are for comparison purpose, and can be improved by EGDCNC, as shown in Fig. \ref{fig:stability_wo_pa} and \ref{fig:stability_w_pa}).

\subsubsection{Effects of Destination Set Size}

Finally, we present in Fig. \ref{fig:tradeoff_2} the delay-cost tradeoffs attained by the GDCNC algorithms, under different destination set sizes varying from $D = 2$ to $5$, and $\lambda = 300$ Mbps.

We make the following observations. A larger destination set size $D$ results in
(i) increasing delay and cost, since more network resources are consumed to handle copies for additional destinations, which also results in longer time waiting for available network resources,
(ii) growing running time, due to the greater number of duplication choices, e.g., GDCNC ($3.2$~s) $\approx$ GDCNC-R ($2.9$~s) when $D = 3$, and GDCNC ($8.9$~s) $>$ GDCNC-R ($5.8$~s) when $D = 5$, 
(iii) a widening gap between the average delay performance of GDCNC and GDCNC-R (in order to attain a given cost, e.g., $20$),
which validates the benefit of GDCNC-R reducing the queuing system and improving the delay performance (which is observed also in Fig. \ref{fig:stability_wo_pa} and \ref{fig:stability_w_pa}).

Again we emphasize that the selection of parameter $V$ targeting optimal cost performance can lead to excessive delay (tens of seconds when $D = 3$), which is a sub-optimal choice
in practical systems. 
With appropriate $V$ values, GDCNC-R can attain suitable delay-cost performance pairs $ (0.8\,\text{s}, 20)$, $(1.3\,\text{s}, 23)$, $(2\,\text{s}, 27)$ under destination set sizes $D = 3, 4, 5$, respectively.

To sum up, although we cannot provide an analytical bound on the performance loss of GDCNC-R, numerical results validate that it can remain competitive in throughput performance (with negligible performance loss), while striking an even better delay-cost tradeoff than GDCNC.

\section{Conclusions}
\label{sec:conclusion}

We addressed the problem of decentralized control of mixed-cast \ac{agi} services in distributed cloud networks. Under the multicast framework, we characterized the enlarged network stability region and analyzed the benefit of exploiting in-network packet duplication. By extending LDP control to a novel queuing system that allows accommodating the duplication operation and making flow control decisions driven by data packets' current destinations, we designed the first decentralized, throughput- and cost-optimal packet processing, routing, and duplication policy for generalized (multicast) flow control, GDCNC, as well as practical variants targeting reduced complexity, enhanced delay, and extended scenarios. Via numerical experiments, we validated the performance gain attained via effective in-network packet duplication, as well as the benefits of joint processing, routing, and duplication optimization for the efficient delivery of multicast AgI services over distributed cloud networks.

\ifCLASSOPTIONcaptionsoff
  \newpage
\fi

\bibliographystyle{IEEEtran}
\bibliography{IEEE_abrv,AgI}




\clearpage

\appendices

\section{Replication Operation Conditions Do Not Sacrifice Performance}
\label{apdx:assumptions}

{\bf Condition a)} and {\bf b)} are justified in Section \ref{sec:duplication}.

{\bf Duplication:}
For any general policy $\mathscr{P}_1$, which can create (and operate on) multiple copies at each time slot, we construct a policy $\mathscr{P}_2$ as follows: for each multicast packet, $\mathscr{P}_2$ performs replication and transmits the resulting packets along the same paths as in $\mathscr{P}_1$, but creating two copies at each time slot and spending multiple slots to complete a replication producing multiple copies under $\mathscr{P}_1$. In the following, we show that the transmission rate of each link under $\mathscr{P}_2$ is the same as that of $\mathscr{P}_1$.


\blue{
Consider the single commodity setting, and denote by $\varsigma$ a possible route to deliver the multicast packet.
Let $f_{ij}^{(\varsigma)}$ denote the average rate of packets traversing link $(i, j)$ that are associated with route $\varsigma$ under $\mathscr{P}_1$, which satisfies $\sum_{\varsigma} f_{ij}^{(\varsigma)} \leq \bar{C}_{ij}$, where $\bar{C}_{ij} = \avg{C_{ij}(t)}$ denotes the average channel capacity of link $(i, j)$. Define
\begin{align}
    \alpha_{ij}^{(\varsigma)} = f_{ij}^{(\varsigma)} \big/ \bar{C}_{ij}.
\end{align}
Then, under $\mathscr{P}_2$, at each time slot, we use $\alpha_{ij}^{(\varsigma)}$ percentage of the instantaneous link capacity $C_{ij}(t)$ to transmit packets associated with route $\varsigma$, i.e., $\{ C_{ij}^{(\varsigma)}(t) = \alpha_{ij}^{(\varsigma)} C_{ij}(t) : \forall\, \varsigma \}$.
We focus on the transmission of packets associated with route $\varsigma$ over link $(i, j)$:
on one hand, the arrival rate of such packets is $f_{ij}^{(\varsigma)}$ (since each packet is delivered along the same path);
on the other hand, the average transmission rate (or {\em service rate}) of such packets is $\avg{C_{ij}^{(\varsigma)}(t)} = \alpha_{ij}^{(\varsigma)} \, \avg{C_{ij}(t)} = \alpha_{ij}^{(\varsigma)} \, \bar{C}_{ij} = f_{ij}^{(\varsigma)}$.
Therefore, the queue collecting {\em packets waiting to cross the link $(i, j)$ and associated with route $\varsigma$} is stable (due to equal arrival and service rates \cite[Theorem~2.4]{Nee:B10}). This result holds for each route $\varsigma$ and link $(i, j)$, concluding the proof.
}

{\bf Non-consolidation:}
Assume that a general policy $\mathscr{P}_1$ consolidates two copies (of status $q_1$ and $q_2$) at node $i$, and we construct a policy $\mathscr{P}_2$ to eliminate this operation as follows. Suppose under $\mathscr{P}_1$, the {\em ancestor} packet splits into two copies of status $(q_1 + s_1)$ and $(q_2 + s_2)$ at node $j$, sent along different routes, where $s_1$ and $s_2$ denote the possible replication operations after the two packets leave node $j$ and before they rejoin at node $i$. Then in $\mathscr{P}_2$, we can create the two copies with status $(q_1 + q_2 + s_1)$ and $s_2$ at node $j$, and send them along the same routes. The system state remains unchanged after $\mathscr{P}_1$ consolidates the two packets, while it saves the network resources to transmit packet $q_2$ from where it is created to node $i$. Therefore, $\mathscr{P}_2$ can achieve the same performance as $\mathscr{P}_1$, if not better.

\section{Multicast Network Stability Region}
\label{apdx:stability_region}

\subsection{Proof of Theorem \ref{thm:stability_1}}

\subsubsection{Necessity}

We need to show: for $\forall\,\V{\lambda} \in \Lambda$, there exist $\V{f}$ and $\V{\beta}$ satisfying \eqref{eq:characterization_1}.

Consider the cost-optimal {\em multicast stabilizing policy} that is under the arrival $\V{\lambda}$. Let $X_{ij}^{(q, s)}(t)$ denote the number of packets successfully delivered to all destinations by time $t$, that {\em underwent the (forwarding and duplication) operation with forwarding choice $(i, j)$ and duplication choice $(q, s)$,} during the delivery.
It is straightforward to obtain the following relationships:
\begin{subequations}
\begin{gather}
    X_{ij}^{(q, s)}(t) \geq 0,\ \forall \, (i, j), (q, s) \\
    \sum_{j\in\delta_i^-} \sum_{s\in 2^{\bar{q}}} X_{ji}^{(q+s, q)}(t) + \sum_{j\in\delta_i^+} \sum_{s\in 2^{\bar{q}}} X_{ij}^{(q+s, s)}(t) + \sum_{\tau = 1}^t a_i^{(q)}(\tau) \leq \sum_{j\in\delta_i^+} \sum_{s\in 2^q} X_{ij}^{(q, s)}(t), \label{eq:conservation} \\
    \sum_{(q, s)} X_{ij}^{(q, s)}(t) \leq t \, C_{ij}, \label{eq:capacity} \\
    X_{d_k \, j}^{(q, s)}(t) = 0,\ \forall \, q: q_k = 1
\end{gather}
\end{subequations}
where \eqref{eq:conservation} is \ac{wrt} status $q$ packets incoming to and outgoing from node $i$, which is an inequality because {\em not all} arrival packets (last term on the left-hand-side) are delivered by time $t$.

\blue{
Divide the above relationships by $t$, let $t \to \infty$, and define $\lim_{t\to\infty} X_{ij}^{(q, s)}(t) \big/ t \triangleq f_{ij}^{(q, s)}$, and we can obtain \eqref{eq:characterization_1}.
In particular, the link capacity constraint \eqref{eq:capacity} becomes:
\begin{align}
\label{eq:link_capacity}
    \lim_{t\to\infty} \Big[ \sum_{(q, s)} \frac{X_{ij}^{(q, s)}(t)}{t} \Big]
    = \sum_{(q, s)} \lim_{t\to\infty} \Big[ \frac{X_{ij}^{(q, s)}(t)}{t} \Big]
    = \sum_{(q, s)} f_{ij}^{(q, s)} \leq C_{ij},
\end{align}
and we define:
\begin{align}
\label{eq:prob_def}
    \beta_{ij}^{(q, s)}
    \triangleq f_{ij}^{(q, s)} \big/ C_{ij}
    \geq 0, \quad \forall\, (i, j) \in \Set{E}, (q, s) \in \Omega.
\end{align}
Then, according to \eqref{eq:link_capacity}: for $\forall\, (i, j) \in \Set{E}$, $\sum_{(q, s)} \beta_{ij}^{(q, s)} \leq 1$, i.e., $\big\{ \beta_{ij}^{(q, s)} : (q, s) \in \Omega \big\}$ is a set of probability values.
In addition, by \eqref{eq:prob_def}, the flow variables satisfy:
\begin{align}
    f_{ij}^{(q, s)} = \beta_{ij}^{(q, s)} \, C_{ij},\quad \forall \, (i, j) \in \Set{E}, (q, s) \in \Omega
\end{align}
which satisfies \eqref{eq:c1_2}, i.e., $f_{ij}^{(q, s)} \leq \beta_{ij}^{(q, s)} \, C_{ij}$.
}

\subsubsection{Sufficiency}
\label{apdx:reference_policy}

We need to show: if there exist $\V{f}$, $\V{\beta}$, $\V{\lambda}$ satisfying \eqref{eq:characterization_1}, then $\V{\lambda} \in \Lambda$.

Consider the stationary randomized policy defined in Theorem \ref{thm:stability_1} (using $\beta_{ij}^{(q, s)} \triangleq f_{ij}^{(q, s)} / C_{ij}$), and denote by $\mu_{ij}^{* (q, s)}(t)$ the associated decisions; then, $\mathbb{E} \big\{ \mu_{ij}^{* (q, s)}(t) \big\} = \big( f_{ij}^{(q, s)} \big/ C_{ij} \big) \, C_{ij} = f_{ij}^{(q, s)}$. Substitute it to \eqref{eq:c1_1}, and we obtain
\begin{align}
    \E{ \mu_{i\to}^{* (q)}(t) - \mu_{\to i}^{* (q)}(t) - \lambda_i^{(q)} } \geq 0
    \ \Longleftrightarrow \ 
    \exists\, \epsilon\geq 0: \E{ \mu_{\to i}^{* (q)}(t) + \lambda_i^{(q)} - \mu_{i\to}^{* (q)}(t)} \leq  - \epsilon
\end{align}
where $\mu_{i \to}^{* (q)}(t)$ and $\mu_{\to i}^{* (q)}(t)$ are defined in \eqref{eq:total_flow}. Furthermore, in \eqref{eq:ldp_bound_1} (in Appendix \ref{apdx:ldp}), we show
\begin{align}
    \E{ \Delta \big( Q_i^{(q)}(t) \big) }
    \leq B + \E{ \big( \mu_{\to i}^{* (q)}(t) + \lambda_i^{(q)} - \mu_{i\to}^{* (q)}(t) \big) Q_i^{(q)}(t) }
    \leq  B - \epsilon \, \E{ Q_i^{(q)}(t) },
\end{align}
which implies the stability of $Q_i^{(q)}(t)$ \cite[Section 3.1.4]{Nee:B10}, and thus $\V{\lambda} \in \Lambda$.

\subsection{Proof of Proposition \ref{prop:enlarged}}

\subsubsection{Unicast Stability Region}
\label{apdx:unicast_nsr}

The unicast stability region is characterized in \cite[Theorem 1]{FenLloTulMol:J18a}, and we rephrase it as follows.

{\em An arrival vector $\V{\lambda}$ is within the stability region $\Lambda_0$ if and only if there exist flow variables $\hat{\V{f}} = \big\{ \hat{f}_{ij}^{(k)} \geq 0 \big\}$ and probability values $\hat{\V{\beta}} = \big\{ \hat{\beta}_{ij}^{(k)} \geq 0: \sum_{k=1}^D \hat{\beta}_{ij}^{(k)} \leq 1 \big\}$ such that:
\begin{subequations}
\label{eq:unicast}
\begin{gather}
    \hat{f}_{\to i}^{(k)} + \lambda_i \leq \hat{f}_{i \to}^{(k)},\quad \forall\, i, k, \label{eq:unicast_conservation} \\
    \hat{f}_{ij}^{(k)} \leq \hat{\beta}_{ij}^{(k)} C_{ij}, \quad \forall\, (i, j),\, k \label{eq:unicast_capacity} \\
    \hat{f}_{d_k \rightarrow}^{(k)}(t) = 0,\quad \forall\, k \label{eq:unicast_boundary}
\end{gather}
\end{subequations}
where $\hat{f}_{\to i}^{(k)} = \sum_{j\in \delta_i^-} \hat{f}_{ji}^{(k)}$ and $\hat{f}_{i \to}^{(k)} = \sum_{j\in \delta_i^+} \hat{f}_{ij}^{(k)}$} (throughout this section, we use the subscripts ``$\to i$\,'' and ``\,$i \to$'' to denote ``$\sum_{j \in \delta_i^-}$'' and ``$\sum_{j \in \delta_i^+}$'' operations on corresponding quantities).

\subsubsection{Proof for ``$\Lambda_0 \subset \Lambda$''}

We aim to show: for any $\V{\lambda}$, $\hat{\V{f}}$, $\hat{\V{\beta}}$ satisfying \eqref{eq:unicast}, the flow variable $\V{f}$ defined as follows, together with the probability values $\beta_{ij}^{(q, s)} = f_{ij}^{(q, s)} \big/ C_{ij}$ and $\V{\lambda}$, satisfies \eqref{eq:characterization_1}:
\begin{align*}
    f_{ij}^{(q, s)} = \sum_{k=1}^D \big( \hat{f}_{ij}^{(k)} / \hat{f}_{i \to}^{(k)} \big) \big[ \lambda_i \, \mathbb{I} \{ (q, s) = ( \V{1} - b_1 - \cdots - b_{k-1}, b_k ) \} + \hat{f}_{\to i}^{(k)} \, \mathbb{I} \{ (q, s) = (b_k, b_k) \} \big].
\end{align*}

The flow variable defined above describes the unicast approach that {\em ``creates one copy for each destination of a multicast packet upon arrival, and treats them as individual unicast packets''}, thus satisfying \eqref{eq:c1_1}.

For any link $(i, j)$, the associated transmission rate is given by 
\begin{align}
    f_{ij} = \sum_{(q, s) \,\in\, \Omega} f_{ij}^{(q, s)} = \sum_{k=1}^{D} (\hat{f}_{ij}^{(k)} / \hat{f}_{i \to}^{(k)}) (\lambda_i + \hat{f}_{\to i}^{(k)}) = \sum_{k=1}^{D} \hat{f}_{ij}^{(k)} = \hat{f}_{ij},
\end{align}
thus satisfying \eqref{eq:c1_2}. In addition, it also indicates that the policies has a cost performance of $h_0^\star(\V{\lambda})$. Therefore, the optimal cost under the multicast framework $h^\star(\V{\lambda}) \leq h_0^\star(\V{\lambda})$.

\subsubsection{Proof for ``$\Lambda \subset D\Lambda_0$''}

Next, we show that: for any $\V{\lambda}$, $\V{f}$, $\V{\beta}$ satisfying \eqref{eq:characterization_1}, the following flow variable $\hat{\V{f}}$, together with probability values $\hat{\beta}_{ij}^{(k)} = \hat{f}_{ij}^{(k)} \big/ C_{ij}$ and $\V{\lambda}' = (\V{\lambda}/D)$, satisfies \eqref{eq:unicast}:
\begin{align}
    \hat{f}_{ij}^{(k)} = \frac{1}{D} \sum_{ (q', s') \,\in\,  \Set{S}_1 } f_{ij}^{(q', s')},
\end{align}
in which
\begin{subequations} \label{eq:qs_def} \begin{gather}
    \Set{S}_1 = \{ (q+s, q): q_k = 1, s\in 2^{\bar{q}} \}, \\
    \Set{S}_2 = \{ (q, s): q_k = 1, s\in 2^{q}\}, \\
    \Set{S}_3 = \{ (q+s, s): q_k = 1, s\in 2^{\bar{q}} \}.
\end{gather} \end{subequations}

To verify the unicast flow conservation law \eqref{eq:unicast_conservation}, we note that:
\begin{align*}
    \hat{f}_{\to i}^{(k)} + \lambda_i'
    = \sum_{ (q', s') \in \Set{S}_1 } \frac{ f_{\to i}^{(q', s')} }{D} + \frac{\lambda_i}{D}
    \overset{\text{(a)}}{\leq} \sum_{ (q', s') \in \Set{S}_2 } \frac{ f_{i \to}^{(q', s')} }{D}
    - \sum_{ (q', s') \in \Set{S}_3 } \frac{ f_{i \to}^{(q', s')} }{D}
    \overset{\text{(b)}}{=} \sum_{ (q', s') \in \Set{S}_1 } \frac{ f_{i \to}^{(q', s')} }{D} = \hat{f}_{i \to}^{(k)}
\end{align*}
where (a) is obtained by first summing up \eqref{eq:c1_1} over $\{ q: q_k = 1\}$
\begin{align}
    \sum_{\{q: q_k = 1\}} \Big[ \, \sum_{s\in 2^{\bar{q}}} f_{\to i}^{(q+s, q)} + \sum_{s\in 2^{\bar{q}}} f_{i \to}^{(q+s, s)} + \lambda_i^{(q)} \Big]
    \leq \sum_{\{q: q_k = 1\}} \sum_{s\in 2^q} f_{i \to}^{(q, s)},
\end{align}
and then plugging in the definition \eqref{eq:qs_def}:
\begin{align}
    \sum_{ (q', s') \in \Set{S}_1 } f_{\to i}^{(q', s')} + \sum_{ (q', s') \in \Set{S}_3 } f_{i \to}^{(q', s')} + \lambda_i
    \leq \sum_{ (q', s') \in \Set{S}_2 } f_{i \to}^{(q', s')},
\end{align}
and (b) results from $\Set{S}_1 = \Set{S}_2 \setminus \Set{S}_3$ as proved in Lemma \ref{lemma:qs}.

For any link $(i, j)$, the associated transmission rate is given by
\begin{align}
    \sum_{k=1}^{D} \hat{f}_{ij}^{(k)}
    = \frac{1}{D} \sum_{k=1}^{D} \sum_{ (q', s') \in \Set{S}_1 } f_{ij}^{(q', s')}
    \overset{\text{(c)}}{\leq} \frac{1}{D} \sum_{k=1}^{D} \sum_{ (q, s) \in \Omega } f_{ij}^{(q, s)}
    = \sum_{(q, s)\in \Omega} f_{ij}^{(q, s)} \leq C_{ij}
\end{align}
where (c) is because $\Set{S}_1 \subset \Omega$, and thus \eqref{eq:unicast_capacity} is verified.

\bigskip

\begin{lemma}
\label{lemma:qs}
$\Set{S}_1 = \Set{S}_2 \setminus \Set{S}_3$.
\end{lemma}

\begin{IEEEproof}
We show the following results:
(i) $\Set{S}_1 \cap \Set{S}_3 = \emptyset$, and
(ii) $\Set{S}_{13} \triangleq \Set{S}_1 \cup \Set{S}_3 = \Set{S}_2$.

To show (i), note that $s_k = 1$ if $(q, s) \in \Set{S}_1$, while $s_k = 0$ if $(q, s) \in \Set{S}_3$, and thus $\Set{S}_1 \cap \Set{S}_3 = \emptyset$.
We prove (ii) in two steps:
\begin{enumerate}
    \item[a)] First, we show $\Set{S}_{13}\subseteq \Set{S}_2$. Take any $(q, s)\in \Set{S}_{13}$,
    \begin{itemize}
        \item If $(q, s)\in \Set{S}_1$, i.e., $q = q' + s'$, $s = q'$ with $q'_k = 1$, which satisfies $q_k = 1$ and $s\in 2^q$, and thus $(q, s) \in \Set{S}_2$.
        \item If $(q, s)\in \Set{S}_3$, i.e., $q = q' + s'$, $s = s'$ with $q'_k = 1$, which satisfies $q_k = 1$ and $s\in 2^q$, and thus $(q, s) \in \Set{S}_2$.
    \end{itemize}
    \item[b)] Next, we show $\Set{S}_2\subseteq \Set{S}_{13}$. Take any $(q, s)\in \Set{S}_2$ (and by definition $q_k = 1$),
    \begin{itemize}
        \item If $s_k = 1$, then we can represent $(q, s)$ as $(q' + s', q')$ with $q' = s$ and $s' = q-s$, which satisfy $q'_k = s_k = 1$ and $s' = q-s \in 2^{\bar{s}} = 2^{\overline{q'}}$, and thus $(q, s) \in \Set{S}_1 \subset \Set{S}_{13}$.
        \item If $s_k = 0$, then we can represent $(q, s)$ as $(q' + s', s')$ with $q' = q-s$ and $s' = s$, which satisfy $q'_k = q_k - s_k = 1$ and $s' = s \in 2^{\overline{q-s}} = 2^{\overline{q'}}$, and thus $(q, s) \in \Set{S}_3 \subset \Set{S}_{13}$.
    \end{itemize}
\end{enumerate}

Combining (i) and (ii) leads to $\Set{S}_1 = \Set{S}_2 \setminus \Set{S}_3$, concluding the proof.
\end{IEEEproof}

\section{Derivation of LDP Bound}
\label{apdx:ldp}

Square the queuing dynamics \eqref{eq:q_dynamic}:
\begin{subequations} \begin{align}
    \big[ Q_i^{(q)}(t+1) \big]^2
    & \leq \big[ Q_i^{(q)}(t) - \mu_{i \to}^{(q)}(t) \big]^2 + \big[ \mu_{\to i}^{(q)}(t) + a_i^{(q)}(t) \big]^2 + 2 \big[ \mu_{\to i}^{(q)}(t) + a_i^{(q)}(t) \big] Q_i^{(q)}(t) \\
    & = \big[ Q_i^{(q)}(t) \big]^2
    - 2 \big[ \mu_{i \to}^{(q)}(t) - \mu_{\to i}^{(q)}(t) - a_i^{(q)}(t) \big] Q_i^{(q)}(t) \\
    & \hspace{10pt} + \big[ \mu_{i \to}^{(q)}(t) \big]^2 + \big[ \mu_{\to i}^{(q)}(t) + a_i^{(q)}(t) \big]^2. \label{eq:constant_bound}
\end{align} \end{subequations} 

We first study the sum of \eqref{eq:constant_bound} over $q \in \{0, 1\}^D$:
\begin{align}
    \sum_{q \in \{0, 1\}^D} \big[ \mu_{i \to}^{(q)}(t) \big]^2
    = \sum_{q \in \{0, 1\}^D} \big[ \sum_{s \in 2^q} \sum_{j \in \delta_i^+}  \mu_{ij}^{(q, s)}(t) \big]^2
    \leq \big[ \sum_{(q, s) \in \Omega} \sum_{j \in \delta_i^+} \mu_{ij}^{(q, s)}(t) \big]^2
    \leq \big( \sum_{j \in \delta_i^+} C_{ij} \big)^2.
\end{align}
Similarly, we can obtain
\begin{align}
    \sum_{q \in \{0, 1\}^D} \big[ \mu_{\to i}^{(q)}(t) + a_i^{(q)}(t) \big]^2
    \leq \big( \sum_{j \in \delta_i^-} C_{ji} + \sum_{j \in \delta_i^+} C_{ij} + A_{i,\max} \big)^2.
\end{align}
Therefore, we can define $B = ( 2 |\delta_{\max}| C_{\max} + A_{i,\max} )^2 / 2$ as a constant bound on the sum of them, where $|\delta_{\max}|$ is the maximum node degree, $C_{\max}$ is the maximum link capacity, and $A_{i,\max}$ is the maximum arrival at node $i$.

The Lyapunov drift of the entire network is given by
\begin{subequations}
\begin{align} \label{eq:controllable}
    \Delta(\V{Q}(t))
    & \triangleq \sum_{i\in \Set{V}} \sum_{q \in \{0, 1\}^D} \frac{Q_i^{(q)}(t+1)^2 - Q_i^{(q)}(t)^2}{2} \\
    & \leq |\Set{V}| B + \sum_{i\in \Set{V}} \sum_{q \in \{0, 1\}^D} a_i^{(q)}(t) Q_i^{(q)}(t) \label{eq:ldp_bound_1} \\
    & \quad - \sum_{i\in \Set{V}} \sum_{q \in \{0, 1\}^D} \underbrace{ \big[ \sum_{s \in 2^q} \sum_{j\in\delta_i^+} \mu_{ij}^{(q,s)}(t) - \sum_{s \in 2^{\bar{q}}} \sum_{j \in\delta_i^-} \mu_{ji}^{(q+s, q)}(t) - \sum_{s \in 2^{\bar{q}}}  \sum_{j\in\delta_i^+} \mu_{ij}^{(q+s, s)}(t) \big] }_{ = \mu_{i \to}^{(q)}(t) - \mu_{\to i}^{(q)}(t)} Q_i^{(q)}(t) \nonumber \\
    & \overset{\text{(a)}}{=} |\Set{V}| B + \sum_{i\in \Set{V}} \sum_{q \in \{0, 1\}^D} a_i^{(q)}(t) Q_i^{(q)}(t) \nonumber \\
    & \quad - \sum_{(i, j) \in \Set{E}} \sum_{(q, s) \in \Omega} \big[ Q_i^{(q)}(t) - Q_j^{(s)}(t) - Q_i^{(q-s)}(t) \big] \mu_{ij}^{(q, s)}(t)
\end{align}
\end{subequations}
where we rearrange the order of summations in (a). Combining the above result with the operational cost model \eqref{eq:cost_model} leads to the weight $w_{ij}^{(q, s)}(t)$ given by \eqref{eq:weight}.

\section{Proof of Theorem \ref{thm:tradeoff}} \label{apdx:tradeoff}

Using \eqref{eq:ldp_bound_1}, we can derive the following \ac{ldp} bound:
\begin{align}
    \Delta( \V{Q}(t) ) + V h(t)
    & \leq |\Set{V}| B - \sum_{i\in \Set{V}} \sum_{q \in \{0, 1\}^D} \big( \mu_{i \to}^{(q)}(t) - \mu_{\to i}^{(q)}(t) - a_i^{(q)}(t) \big) Q_i^{(q)}(t) + V h(t) \nonumber \\
    & \leq |\Set{V}| B - \sum_{i\in \Set{V}} \sum_{q \in \{0, 1\}^D} \E{ \big( \mu_{i \to}^{* (q)}(t) - \mu_{\to i}^{* (q)}(t) - a_i^{(q)}(t) \big) Q_i^{(q)}(t) } + V \E{ h^*(t) } \nonumber \\
    & \overset{\text{(a)}}{\leq} |\Set{V}| B - \epsilon \sum_{i\in \Set{V}} \sum_{q \in \{0, 1\}^D} \E{ Q_i^{(q)}(t) } + V h^\star( \V{\lambda} + \epsilon \V{1} )
\end{align}
where $\V{\mu}^*(t)$ denotes the decisions associated with the cost-optimal stationary randomized policy under arrival vector $\V{\lambda} + \epsilon \V{1}$ (defined in Theorem \ref{thm:stability_1}).
In (a), note that:
(i) the decisions are independent with $Q_i^{(q)}(t)$, and thus expectation multiplies,
(ii) $\E{ \mu_{i \to}^{* (q)}(t) - \mu_{\to i}^{* (q)}(t) -  ( \lambda_i^{(q)} + \epsilon ) } \geq 0$ because the randomized policy stabilizes $\V{\lambda} + \epsilon \V{1}$,
(iii) the policy is cost-optimal under $\V{\lambda} + \epsilon \V{1}$.

Fix any $T > 0$, and apply telescope sum \cite{Nee:B10} to the interval $[0, T-1]$ (assuming empty queues at $t = 0$). Divide the result by $T$, push $T\to\infty$, and we obtain:
\begin{align}
    V \avg{ \E{h(t)} }
    \leq |\Set{V}| B - \epsilon \avg{ \E{ \| \V{Q}(t) \|_1 } } + V h^\star( \V{\lambda} + \epsilon \V{1} ).
\end{align}

Based on this inequality, we can derive:

(i) {\em Average queue backlog:} note that $\avg{ \E{h(t)} } \geq h^\star(\V{\lambda})$, and we obtain
\begin{align}
    \avg{ \E{ \| \V{Q}(t) \|_1 } }
    \leq \frac{ |\Set{V}| B }{\epsilon} + \left[ \frac{h^\star(\V{\lambda} + \epsilon \V{1}) - h^\star(\V{\lambda})}{\epsilon} \right] \, V.
\end{align}

(ii) {\em Operational cost:} dropping the second term in the right hand side, we obtain
\begin{align}
    \avg{ \E{h(t)} }
    \leq h^\star(\V{\lambda} + \epsilon \V{1}) + \frac{ |\Set{V}| B }{V}\quad 
    \xrightarrow[]{\ \{\epsilon_n \} \, \downarrow \, 0 \ }\quad 
    \avg{ \E{h(t)} }
    \leq h^\star(\V{\lambda}) + \frac{|\Set{V}| B}{V}
\end{align}
where we take a sequence $\{\epsilon_n \} \, \downarrow \, 0$ and use the fact that the inequality holds for any $\epsilon_n > 0$.

\section{Average Queuing Delay}
\label{apdx:delay}

Each duplication tree $\mathcal{T}$ is a possible way to accomplish the goal of multicast packet delivery. We divide the physical queue $Q^{(q)}(t) = \sum_{i\in \Set{V}} Q_i^{(q)}(t)$ into sub-queues $Q_\Set{T}^{(q)}(t)$ associated with each duplication tree $\mathcal{T}$, and $Q^{(q)}(t) = \sum_{ \Set{T} \in \, \Set{U}(q) } Q_\Set{T}^{(q)}(t)$ where $\Set{U}(q) = \{ \Set{T} : q\in \Set{T} \}$ denotes the set of duplication trees including status $q$ as a tree node.

With this model, the average delay can be derived in two steps:
(i) calculate the average delay $\Delta_\Set{T}$ for each duplication tree $\Set{T}$,
(ii) calculate the average delay of all trees (weighted by $\lambda_\Set{T}$, which is the rate of packets selecting each duplication tree $\Set{T}$):

(i) Consider a given duplication tree $\Set{T}$, with an associated packet rate of $\lambda_\Set{T}$. First, we focus on the delivery of the $k$-th copy, i.e., the path from the root node $\V{1}$ to the leaf node $b_k$, denoted by $\omega_k$ (which is composed of all tree nodes in the path). By Little's Theorem, the average delay of this path is given by $\Delta_{\Set{T}}(k) = \sum_{q \in \omega_k} \bar{Q}_\Set{T}^{(q)} / \lambda_{\Set{T}}$, in which $\bar{Q}_\mathcal{T}^{(q)} = \avg{ \E{ Q_\Set{T}^{(q)}(t) } }$, and we average over all the copies $k= 1,\cdots, D$:
\begin{align}
\label{eq:tree_average}
    \Delta_{\Set{T}} 
    = \frac{1}{D} \sum_{k=1}^D \Delta_{\Set{T}}(k)
    = \frac{1}{D} \sum_{k=1}^D \sum_{q\in\omega_k} \frac{\bar{Q}_\Set{T}^{(q)}}{\lambda_{\Set{T}}}
    \overset{\text{(a)}}{=} \frac{1}{D\lambda_{\mathcal{T}}} \sum_{q\in \Set{T}} \|q\|_1 \bar{Q}_\mathcal{T}^{(q)}
\end{align}
where we exchange the order of summations in (a), and use the fact that node $q$ in the duplication tree $\Set{T}$ is included in $\|q\|_1$ different paths (e.g., each leaf node $b_k$ belongs to one path $\omega_k$, the root node $\V{1}$ belongs to all paths $\omega_1, \cdots, \omega_D$).

(ii) With the average delay of each duplication tree given by \eqref{eq:tree_average}, the overall average delay can be derived as follows
\begin{align} \begin{split}
    \Delta
    & = \sum_{\Set{T}} \frac{ \lambda_{\Set{T}} }{ \|\V{\lambda}\|_1 } \Delta_{\Set{T}}
    = \sum_{\Set{T}} \frac{ \lambda_{\Set{T}} }{ \|\V{\lambda}\|_1 } \Big[ \frac{1}{D\lambda_{\mathcal{T}}} \sum_{q\in \Set{T}} \|q\|_1 \bar{Q}_\mathcal{T}^{(q)} \Big]
    = \frac{1}{D\|\V{\lambda}\|_1} \sum_{\Set{T}} \sum_{q\in \Set{T}} \|q\|_1 \bar{Q}_\mathcal{T}^{(q)} \\
    & \overset{\text{(b)}}{=} \frac{1}{D\|\V{\lambda}\|_1} \sum_{q \in \{0, 1\}^D} \|q\|_1 \sum_{\Set{T}\in \, \Set{U}(q)} \bar{Q}_\mathcal{T}^{(q)}
    = \frac{1}{D\|\V{\lambda}\|_1} \sum_{q \in \{0, 1\}^D} \|q\|_1 \bar{Q}^{(q)} \\
    & = \frac{1}{D\|\V{\lambda}\|_1} \sum_{q \in \{0, 1\}^D} \|q\|_1 \avg{\E{Q^{(q)}(t)}}
\end{split} \end{align}
where we exchange the order of summations in (b). Therefore, the average delay is linear in the queue backlog, and the coefficient of the status $q$ queue is proportional to $\|q\|_1$.

\section{Transformation of AgI Service Delivery into Packet Routing}
\label{apdx:AgI}

\subsection{Constructing the Layered Graph}

Denote the topology of the actual network by $\Set{G} = (\Set{V}, \Set{E})$. For any service $\phi$ (consisting of $M_{\phi} - 1$ functions), the layered graph $\Set{G}^{(\phi)}$ is constructed as follows:
\begin{itemize}
	\item[(i)] make $M_{\phi}$ copies of the actual network, indexed as layer $1,\cdots, M_{\phi}$ from top to bottom, and we denote node $i\in \Set{V}$ on the $m$-th layer by $i_m$;
	\item[(ii)] add {\em directed} links connecting corresponding nodes between adjacent layers, i.e., $(i_m, i_{m+1})$ for $\forall\, i\in \Set{V}$, $m = 1,\cdots, M_{\phi} - 1$.
\end{itemize}
To sum up, the layered graph $\Set{G}^{(\phi)} = (\Set{V}^{(\phi)}, \Set{E}^{(\phi)})$ is defined as
\begin{subequations}
\begin{align}
    \Set{V}^{(\phi)} & = \{ i_m : i\in \Set{V}, 1\leq m\leq M_{\phi} \} \\
    \Set{E}^{(\phi)}_{\text{pr}, i} & = \{ (i_m, i_{m+1}) : , 1\leq m\leq M_{\phi}-1 \} \\
    \Set{E}^{(\phi)}_{\text{tr}, ij} & = \{ (i_m, j_m) : (i, j)\in \Set{E}, 1\leq m\leq M_{\phi} \}
\end{align}
\end{subequations}
with $\Set{E}^{(\phi)} = \{ \Set{E}^{(\phi)}_{\text{pr}, i}: i\in \Set{V} \} \cup \{ \Set{E}^{(\phi)}_{\text{tr}, ij}: (i, j)\in \Set{E} \}$.

{\em Physical interpretation:} Layer $m$ in $\Set{G}^{(\phi)}$ accommodates packets of stage $m$. The edges in $\Set{E}^{(\phi)}_{\text{pr}, i}$ and $\Set{E}^{(\phi)}_{\text{tr}, ij}$ indicate the processing and transmission operations in the actual network, respectively. More concretely, the flow on $(i_m, i_{m+1})$ denotes the processing of stage $m$ packets by function $m$ at node $i$, while $(i_m, j_m)$ denotes the transmission of stage $m$ packets over link $(i, j)$. In addition, $\Set{D}^{(\phi)}_{M_{\phi}}$ denotes the destination set in the graph (since only stage $M_{\phi}$ packets are consumable), where $\Set{D}^{(\phi)}$ is the destination set of service $\phi$ in the actual network.

We define two parameters $(\zeta_{\imath\jmath}^{(\phi)}, \rho_{\imath\jmath}^{(\phi)})$ for each link $(\imath, \jmath)$ in the layered graph
\begin{align}
    (\zeta_{\imath\jmath}^{(\phi)}, \rho^{(\phi)}_{\imath\jmath})
    = \begin{cases}
    (\xi_{\phi}^{(m)}, r_{\phi}^{(m)}) & (\imath, \jmath) = (i_m, i_{m+1}) \\
    (1, 1) & (\imath, \jmath) = (i_m, j_m)
    \end{cases}.
\end{align}
The two parameters can be interpreted as the generalized scaling factor and workload: for transmission edges (the second case), $\zeta_{\imath\jmath}^{(\phi)} = 1$ since flow is neither expanded or compressed by the transmission operation, and $\rho_{\imath\jmath}^{(\phi)} = 1$ since the flow and the transmission capability are quantified based on the same unit.

\subsection{Relevant Quantities}

We define flow variable $\mu_{\imath\jmath}^{(\phi, q, s)}(t)$ for link $(\imath, \jmath)$ in the layered graph, which is the flow sent to the corresponding interface, leading to the received flow by node $\jmath$ given by $\zeta_{\imath\jmath}^{(\phi)} \mu_{\imath\jmath}^{(\phi, q, s)}(t)$. By this definition, for $\forall \, \phi$ and $\imath \in \Set{G}^{(\phi)}$, the queuing dynamics are modified by
\begin{align}\begin{split}
Q_\imath^{(\phi, q)}(t+1)
\leq \max \big[ Q_\imath^{(\phi, q)}(t) 
- \mu_{\imath\to}^{(\phi, q)}(t), 0 \big]
+ \mu_{\to\imath}^{(\phi, q)}(t)
+ a_\imath^{(\phi, q)}(t)
\end{split}\end{align}
where the incoming and outgoing flows are given by
\begin{subequations} \begin{align}
    \mu_{\imath\to}^{(\phi, q)}(t) & = \sum_{\jmath\in \delta_\imath^+} \sum_{s\in 2^{q}} \mu_{\imath\jmath}^{(\phi, q, s)}(t), \\  
    \mu_{\to\imath}^{(\phi, q)}(t) & = \sum_{\jmath\in \delta_\imath^-} \sum_{s\in 2^{\bar{q}}} \zeta_{\jmath\imath}^{(\phi)} \mu_{\jmath\imath}^{(\phi, q+s, q)}(t) + \sum_{\jmath\in \delta_\imath^+} \sum_{s\in 2^{\bar{q}}} \mu_{\imath\jmath}^{(\phi, q+s, s)}(t).
\end{align} \end{subequations}

The modified link capacity constraints are
\begin{subequations}
\begin{align} \label{eq:agi_resource}
\sum_{(\imath, \jmath)\in \Set{E}_{\text{pr}, i}^{(\phi)}} \sum_{\phi} \sum_{(q, s)\in \Omega} \rho_{\imath \jmath}^{(\phi)} \mu_{\imath \jmath}^{(\phi, q, s)}(t) & \leq C_{i}, \\ 
\sum_{(\imath, \jmath)\in \Set{E}_{\text{tr}, ij}^{(\phi)}} \sum_{\phi} \sum_{(q, s)\in \Omega} \rho_{\imath \jmath}^{(\phi)} \mu_{\imath \jmath}^{(\phi, q, s)}(t) & \leq C_{ij},
\end{align}
\end{subequations}
and the modified resource operational cost is
\begin{align}
h(t) = \sum_{(\imath, \jmath)\in \Set{E}^{(\phi)}} e_{\imath \jmath} \sum_{\phi} \sum_{(q, s)\in \Omega} \rho_{\imath \jmath}^{(\phi)} \mu_{\imath \jmath}^{(\phi, q, s)}(t)
\end{align}
where $e_{i_m i_{m+1}} = e_i$ and $e_{i_m j_m} = e_{ij}$.

Similarly, the goal is to minimize the time average cost while stabilizing the modified queues, and we can follow the procedure in Section \ref{sec:ldp}, i.e., deriving the upper bound for \ac{ldp} and formulating an optimization problem to minimize the bound, and the derived solution is in the max-weight fashion as shown in Section \ref{sec:control_alg} and \ref{sec:AgI}.

\begin{remark}
In contrast to data transmission, the processing operation can expand/compress data stream size, and queues of expanding data streams can attract more attention in the developed algorithm. To address this problem, we can normalize the queues by $\tilde{Q}_\imath^{(\phi, q)}(t) = Q_\imath^{(\phi, q)}(t) \big/ \Xi_{\phi}^{(m)}$ if $\imath = i_m$, in which
\begin{align}
    \Xi_{\phi}^{(m)} = \prod_{s=1}^{m-1} \xi_{\phi}^{(m)},\text{ and }
    \Xi_{\phi}^{(1)} = 1
\end{align}
is interpreted as the {\em cumulative scaling factor} till stage $m$. The resulting design, i.e., optimize the drift of the normalized queues, can achieve a better balance among the queues, while preserving (throughput and cost) optimality.
\end{remark}

\section{Complexity Analysis}
\label{apdx:complexity}

\subsection{GDCNC}

The number of all duplication choices $(q, s)$ is given by
\begin{align}
|\Omega| = \sum_{k=1}^{D} \operatorname{C}(k; D) (2^k - 1) = 3^D - 2^D \sim \mathcal{O}(3^D)
\end{align}
where $\operatorname{C}(k; D)$ denotes the combinatorial number of {\em choosing $k$ from $D$}.
To wit, we divide the elements of $\Omega$, $(q, s)$, into $D$ groups based on the first element $q$: group $k\ (k = 1, \cdots, D)$ collects the $(q, s)$ pairs such that $q$ has $k$ entries equal to $1$, including $\operatorname{C}(k; D)$ different $q$; in addition, fix $q$, there are $|2^{q}| - 1 = 2^k - 1$ different $s$ (other than $\V{0}$).
Therefore, the complexity of GDCNC, which is proportional to $|\Omega|$, is $\mathcal{O}(3^D)$.

\subsection{GDCNC-R}

The key problem is to calculate the number of nodes in each duplication tree.

We note that each duplication tree includes $D-1$ duplication operations (which can be shown by mathematical induction); in addition, each duplication operation involves $3$ nodes, i.e., the parent node $q$ and two child node $s$ and $r$, and the total number of nodes is $3(D-1)$.
However, note that every node is counted {\em twice} -- one time as the parent $q$ (packet to duplicate), and the other time as a child $s$ or $r$ (the created copy) -- other than the root node and the $D$ leaf nodes, since there is no duplication operation with the root node $\V{1}$ as a child, or a leaf node $b_k$ as the parent, i.e., these $D+1$ nodes are counted only {\em once}.

As a result, the number of internal nodes other than the root node is:
\begin{align}
\frac{3(D-1) - (D+1)}{2} = D - 2,
\end{align}
and together with the root node, each tree has $D-1 = (D-2) + 1$ internal nodes (and each of them is associated with $3$ duplication choices), and $D$ leaf nodes (and each of them is associated with $1$ duplication choice), leading to a complexity of $\mathcal{O}(D)$.

\section{Notes on Generalized Flow Conservation Constraint} \label{apdx:flow_conservation}

In \cite[Eq (4)]{cai2021multicast}, we present the following flow conservation law:
\begin{align} \label{eq:flow_conservation_a}
    \sum_{\{ q: q_k = 1 \}} \big[ \breve{f}_{\to i}^{(q)} + \lambda_i^{(q)} \big]
    = \sum_{\{ q: q_k = 1 \}} \breve{f}_{i \to}^{(q)}
\end{align}
where $\breve{f}_{\to i}^{(q)}$ and $\breve{f}_{i \to}^{(q)}$ are the total incoming/outgoing flow rates of status $q$ packets to/from node $i$, given by
\begin{align} \label{eq:link_flow}
    \breve{f}_{\to i}^{(q)} = \sum_{s\in 2^{\bar{q}}} \sum_{j\in \delta_i^-} f_{ji}^{(q+s, q)},\ 
    \breve{f}_{i \to}^{(q)} = \sum_{s\in 2^{\bar{q}}} \sum_{j\in \delta_i^+} f_{ij}^{(q+s, q)}.
\end{align}

Despite of the neat form and clear insight of this relationship, we will show that the generalized flow conservation constraint \eqref{eq:flow_conservation} is a more fundamental characterization for the in-network packet duplication operation.

\underline{First, we show \eqref{eq:flow_conservation}  $\Rightarrow$ \eqref{eq:flow_conservation_a}}.
Substitute \eqref{eq:link_flow} into \eqref{eq:flow_conservation_a}, and the result to be shown is given by:
\begin{align}  \label{eq:fc1}
\sum_{\{q: q_k = 1\}} \Big[ \, \sum_{s\in 2^{\bar{q}}} \sum_{j\in \delta_i^-} f_{ji}^{(q+s, q)} + \lambda_i^{(q)} \Big]
= \sum_{\{q: q_k = 1\}} \sum_{s\in 2^{\bar{q}}} \sum_{j\in \delta_i^+} f_{ij}^{(q+s, q)}.
\end{align}

Sum up \eqref{eq:c1_1} over $\{ q: q_k = 1\}$, and we obtain
\begin{align} \label{eq:fc2}
    \sum_{\{q: q_k = 1\}} \Big[ \, \sum_{s\in 2^{\bar{q}}} \sum_{j\in \delta_i^-} f_{ji}^{(q+s, q)} + \sum_{s\in 2^{\bar{q}}} \sum_{j\in \delta_i^+} f_{ij}^{(q+s, s)} + \lambda_i^{(q)} \Big]
    = \sum_{\{q: q_k = 1\}} \sum_{s\in 2^q} \sum_{j\in \delta_i^+} f_{ij}^{(q, s)}.
\end{align}
Compare \eqref{eq:fc1} and \eqref{eq:fc2}, and it remains to be shown that
\begin{align}
    \sum_{j\in \delta_i^+} \sum_{ \{(q, s): q_k = 1, s\in 2^{q}\} } f_{ij}^{(q, s)}
    = \sum_{j\in \delta_i^+} \sum_{ \{(q, s): q_k = 1, s\in 2^{\bar{q}}\} } \big[ f_{ij}^{(q+s, q)} + f_{ij}^{(q+s, s)} \big],
\end{align}
or equivalently,
\begin{align} \label{eq:qs_space}
\sum_{j\in \delta_i^+} \Big[ \sum_{ (q, s)\in \Set{S}_2 } f_{ij}^{(q, s)} - \sum_{ (q, s)\in \Set{S}_1 } f_{ij}^{(q, s)} - \sum_{ (q, s)\in \Set{S}_3 } f_{ij}^{(q, s)} \Big] = 0
\end{align}
where $\Set{S}_1$, $\Set{S}_2$, $\Set{S}_3$ are defined in \eqref{eq:qs_def} and satisfy $\Set{S}_1 = \Set{S}_2 \setminus \Set{S}_3$ as shown in Lemma \ref{lemma:qs}. Therefore, each term in the summation (over $j\in \delta_i^+$) equals $0$.

Next, we present a counterexample to \underline{show \eqref{eq:flow_conservation_a} is not sufficient} to guarantee the existence of a feasible policy:
Consider $D= 3$ destinations. Let $\breve{f}_{\to i}^{(q)} = 1$ for $q = \V{1} = (1, 1, 1)$ and $q = b_1 = (1, 0, 0)$, $\breve{f}_{i \to}^{(q)} = 1$ for $q = (1, 1, 0)$ and $q = (1, 0, 1)$, and the other flow variables and $\lambda_i^{(q)}$ be $0$.
This flow assignment satisfies \eqref{eq:flow_conservation_a}.

The corresponding operation is described as follows: (i) first ``consolidate'' the incoming packets of $(1, 1, 1)$ and $(1, 0, 0)$, and (ii) then create (by duplication) outgoing packets of $(1, 1, 0)$ and $(1, 0, 1)$. However, the operation in step (i) cannot be realized in the actual network, because the packets are not of identical content (since the status $\V{1}$ packet is prior to any duplication, whose content is different from the status $b_1$ packet) and cannot be consolidated.
In fact, there does not exist flow variable satisfying \eqref{eq:flow_conservation} to describe the above operation.


\end{document}